\documentclass[aps,twocolumn,showpacs,amsmath,amssymb,pra,superscriptaddress,floatfix]{revtex4-1}
\usepackage{braket}
\usepackage{amsmath}
\usepackage{amssymb}
\usepackage{mathtools}
\usepackage{graphicx}
\usepackage{dcolumn}
\usepackage{bm}
\usepackage{multirow}
\usepackage{leftidx}
\usepackage{color}
\usepackage{float}

\usepackage{amsfonts}
\usepackage{eufrak}
\usepackage[german,english]{babel}

\begin{document}
\title{Schwinger variational principle theory of collisions in the presence of multiple potentials}

\author{F. Robicheaux}
\email{robichf@purdue.edu}
\affiliation{Department of Physics and Astronomy, Purdue University, West Lafayette,
Indiana 47907, USA}

\author{P. Giannakeas}
\email{pgiannak@purdue.edu}
\affiliation{Department of Physics and Astronomy, Purdue University, West Lafayette, Indiana 47907, USA}

\author{Chris H. Greene}
\email{chgreene@purdue.edu}
\affiliation{Department of Physics and Astronomy, Purdue University, West Lafayette, Indiana 47907, USA}

 \date{\today}

\begin{abstract}
A theoretical method for treating collisions in the presence of
multiple potentials
is developed by employing the Schwinger variational principle.
The current treatment agrees with the local
(regularized) frame transformation theory and extends its capabilities. Specifically,
the Schwinger variational approach gives results without the
divergences that need to be regularized in other methods.
Furthermore, it provides a framework to
identify the origin of these singularities and possibly
improve the local frame transformation. We have used the method
to obtain the scattering parameters for different confining
potentials symmetric in $x,y$.
The method is also used to treat photodetachment processes in the
presence of various confining potentials, thereby highlighting 
effects of the infinitely many closed channels.
Two general features predicted are the vanishing of the total photoabsorption probability 
at {\it every} channel threshold and the occurrence of resonances below the channel thresholds for negative scattering lengths.
In addition, the case of negative ion photodetachment in the
presence of uniform magnetic fields is also considered where
unique features emerge at large scattering lengths.
\end{abstract}


\maketitle

\section{Introduction}

Frame transformation theory constitutes an important theoretical
toolkit describing Hamiltonian systems with several potential
terms which are dominant in different regions.
A particular class of frame transformations, i.e. the local
frame transformation (LFT), can provide
pivotal insights for quantum mechanical systems which are
approximately separable, usually in different coordinates,
at small or large distances but are not separable over all space.
In their seminal papers, Fano\cite{fano1981stark} and
Harmin\cite{harminpra1982,harminprl1982,harmin1981hydrogenic}
developed the principles of the LFT
theory permitting the theoretical treatment of highly
excited alkali atoms in the presence of external electric fields
and interpreting on physical grounds the corresponding
photoabsorption spectra.

In addition, the unique properties and intuitive insights make
the LFT theory the key framework in the investigations of
atom-electron collisions under the action of uniform
magnetic\cite{Greene1987pra} or electric
fields\cite{WongRauGreene1988pra,RauWong1988pra,GreeneRouze1988ZPhys,SlonimGreene1991}.
The description of the corresponding photoabsorption spectra in
terms of the LFT approach is accurate even though only a few
channels were involved in the actual calculations.

However, in 1966 it was shown that the inclusion of
{\em all} closed channels yields non-trivial effects 
when a charged particle is scattered by a zero-range potential
in the presence of a magnetic field \cite{demkovjetp1966}.
In particular, it was shown that there always exists a bound state for negative scattering length
that is collectively supported by all the closed channels (Landau levels)
induced by the magnetic field.
The properties of this bound state were subsequently investigated in the context of negative-ion photodetachment in weak magnetic fields \cite{grozdanovpra1995}.
However, similar effects also appear in ultracold two-body collisions
in tight waveguide geometries yielding confinement-induced
resonances of the Fano-Feshbach type\cite{olshanii1998prl}.
Such scenarios have been studied in terms of LFT theory either
for atom-atom
scattering\cite{GrangerBlume2004prl,Giannakeas2012pra,hess2014pra,Zhang2013pra}
or dipole-dipole collisions\cite{giannakeas2013prl}.
LFT theory provided the most complete theoretical framework in order to describe the relevant collisional physics encompassing the concept of infinitely many closed channels.
Interestingly, this permitted the theory to go beyond previous studies where the impact of the closed channels is included in an averaging sense \cite{kim2005,crawfordpra1988}.
However, the integration of the concept of infinitely many
closed channels in the LFT approach yielded divergences, which required
regularization techniques to be employed to remove the singular
behavior\cite{GrangerBlume2004prl,Giannakeas2012pra,Zhang2013pra,giannakeas2013prl}.
Evidently, an approach that incorporates the concept of
infinitely many closed channels while avoiding the infinities
of the LFT treatment could be useful.
Such a formalism could provide quantitative
insights into the origins of the LFT singularities while preserving
the physical intuition of LFT.
Also, it could allow the systematic study of non-trivial effects
arising from the full Hilbert space in collisional systems, e.g.
the scattering of an electron in a Rydberg state from
a neutral perturber\cite{dupra1987}.

In this paper, a Schwinger variational $K$-matrix
approach\cite{schwinger1947,schwinger1950,schwingerlippmann1950} is
developed for treating Hamiltonians with two potential terms
which dominate at different length scales. We 
restrict our treatment in this paper
to the situation of symmetric confining potentials
in $x,y$ and no force in $z$; the more general case of asymmetric
potentials and/or forces in $z$ will be treated elsewhere.
Note that the Schwinger variational principle has been
successfully used for the quantitative study of Rydberg
molecules\cite{stephens1992jcp}, Rydberg atoms \cite{goforthpra1992}, electron
scattering\cite{lucchese1979jpb,  malekipra1980} and electron-molecule collisions\cite{lucchese1983pra,lino2000bjp,lucchese1983amopp}.
However, in most of these studies, the trial function is composed
from a superposition of basis functions (often Gaussian orbitals),
which requires numerical evaluation of
complicated volume integrals.
In our analysis, such complications are avoided due to the length
scale separation between the two potential terms, permitting us
to treat the scattering information analytically.

This paper will focus on the case of a short-range spherically symmetric
interaction potential with a confining potential that bounds all
the degrees of freedom apart from one at large distances.
Note that these types of Hamiltonians refer to the physical
systems of atom-atom collisions in tight waveguides
or negative ion photodetachment processes in external fields.
Under these considerations, the calculations within
the Schwinger variational framework are compared to the
corresponding {\it regularized} LFT derivations for two different
confining potentials (a harmonic oscillator and an infinite square
well potential), yielding identical results.
However, in the current Schwinger variational
$K$-matrix approach, the infinitely many
closed channels associated with the confining potential produces finite
results without regularization.

Our analysis includes several applications of the current
formalism to collisions in the presence of
either harmonic or infinite square well confining
potentials where the corresponding photoabsorption spectra
for $s$-wave photoelectrons possesses unique features.
Namely, the total photoabsorption probability vanishes
at every channel threshold regardless of
the specific details of the confining geometries. This
threshold phenomenon illustrates the impact of the closed channel
physics on the collisional complex.
In addition to the cases of harmonic and infinite square well
confining potentials, the current development is applied
also to infinite rectangular potentials or off-centered
scattering in an infinite square well.
These particular systems illustrate the
physics of infinitely many closed channels in the
absence of degeneracies in the spectra of the confining Hamiltonians.
As a final application, negative ion
photodetachment in the presence of a uniform magnetic
field is treated.
Note this particular system has been thoroughly investigated both experimentally \cite{larsonpra1985, krausseprl1990} and theoretically \cite{Greene1987pra,larsonpra1979,crawfordpra1988}.
However, in most of the theoretical investigations only the open channel physics was considered and this permits us to explicitly show the impact of the closed channels.
More specifically, our results are compared to the
approximate LFT treatment in Ref.~\cite{Greene1987pra} where
only the open channels physics were considered.
Qualitative changes occur only near thresholds for small scattering lengths.
However, these features are enhanced the cases where the scattering length is comparable
to the confinement length scale which implies enormous
scattering lengths for laboratory strength magnetic fields
or ultrastrong magnetic fields for the usual scattering lengths.

Section II introduces the basic concepts which are used
in the rest of this work.
Section III focuses on the derivation of the Schwinger
variational $K$-matrix for the confining potentials of
a harmonic or a square well potential assuming that the
short-range spherically symmetric potential induces partial
waves with angular and azimuthal quantum numbers
$(\ell m)=[(00),(10),(1\pm1)]$.
In addition, connections between the current formalism
and the LFT theory are discussed.
Section IV is devoted in the discussion of several
applications of the Schwinger variational $K$-matrix
as well as its comparison with numerical simulations.
Several appendices give some of the details of the derivations.
Finally Section V summarizes and concludes our analysis.

\section{Basic definitions and relations}

This section contains the conventions used in this paper for the different Hamiltonians with additive potentials, the
scattering wave functions as well as
the basic scattering parameters (e.g. the $K$-matrix).
In addition, the definitions of the corresponding Green's functions are also included.
Since there are many possible conventions, this will allow for unambiguous definitions of parameters.

\subsection{Hamiltonians}

The scope of this work is to treat the interaction of a particle with different
potentials that have qualitatively different character, with the
focus on the case where the particle's motion is unbounded in
one direction.
The continuum solutions can be obtained and used for the
calculation of scattering parameters, photoionization and/or
photodetachment cross sections, etc. 
The full scattering information is encapsulated in the full Hamiltonian $\hat{H}$ which reads:
\begin{equation}
\hat{H} \ket{\psi} = (\hat{H}_f + \hat{V})\ket{\psi}= E\ket{\psi},
\label{eq1}
\end{equation}
where $\hat{H}_f=p^2/2\mu$ is the free particle Hamiltonian and $\mu$ indicates its mass.
$\hat{V}$ corresponds to the operator of the total potential,
$\ket{\psi}$ is the scattering wave function and
$E$ is the total energy.

In the following we assume that the operator $\hat{V}$
in the full Hamiltonian
$\hat{H}$ is the sum of two potentials $\hat{V}_s$ and
$\hat{V}_c$, which possess simple functional forms
in different coordinate systems. The two potentials are such that
the wave function is easily calculated for each potential individually.
More specifically, the $\hat{V}_c$ will be a smooth potential
that extends over large separation distances and affects the
asymptotic boundary conditions in $\ket{\psi}$.
In this study, the $\hat{V}_c$ is considered
to bound the motion of the
particle in all the degrees of freedom except one.
For example,
this potential could be a harmonic one,
$V_c(\mathbf{r})=(\mu /2)\omega_\perp^2 (x^2+y^2)$ with frequency $\omega_\perp$, 
which separates in cylindrical coordinates, or an infinite
square well on the $x-y$ plane which separates in
Cartesian coordinates.
The $V_s(\mathbf{r})$ is a potential that is non-zero in a
small volume $\tau$ and could be a spherically
symmetric potential at the origin.

One key observation concerns the length scale separation
between the $\hat{V}_s$ and $\hat{V}_c$ potential terms.
With this assumption in mind two new Hamiltonians are defined
which use solely one potential term each.
The $\hat{H}_c$ is the Hamiltonian that only contains
$\hat{V}_c$ and does {\it not} contain the $V_s$.
\begin{equation}
\hat{H}_c\ket{\psi_c} = E\ket{\psi_c},
\label{eq2}
\end{equation}
where $\ket{\psi_c}$ is the corresponding wave function
and $E$ is the total energy.
The solutions are analytic functions and are used
in the following to define the asymptotic form of the
scattering wave function.
In addition, the corresponding collisional information
can be expressed in terms of these states.

Consider now the Hamiltonian $\hat{H}_s$ which contains only the $\hat{V}_s$ potential.
Then the corresponding Schr\"odinger equation reads

\begin{equation}
\hat{H}_s\ket{\psi_s} = E\ket{\psi_s},
\label{eq3}
\end{equation}
where the eigenfunctions of this Hamiltonian have a simple
form because they do not include the complicated boundary
conditions that arise from the confining potential, $\hat{V}_c$.
Intuitively, Eq.~(\ref{eq3}) describes the motion of the particle at small distances where it experiences solely the short-range potential $\hat{V}_s$.

\subsection{Scattering wave functions and $K$-matrix}
In this subsection the scattering wave functions of the $\hat{H}_s$ and $\hat{H}_c$ Hamiltonians are defined.
More specifically, for the Hamiltonian $\hat{H}_s$ the corresponding potential term, namely $\hat{V}_s$, possesses spherical symmetry and is short-ranged. Therefore, the $\ket{\psi_s}$ wave functions in Eq.~(\ref{eq3}) are expressed in spherical coordinates as radial functions times spherical harmonics. 
At distances $r>r_0$, the $\hat{V}_s$ is negligible, namely $V_s(r_0)\approx 0$, therefore the $K$-matrix normalized radial functions are
phase-shifted from the free particle wave function according to the following relation: 
\begin{eqnarray}\label{eqPsil}
\braket{\mathbf{r}|\psi_s}&=&\psi_{s,\ell m} (\mathbf{r})\cr
&=& \sqrt{\frac{2\mu k}{\pi\hbar^2}}[j_\ell (kr)-
 \tan\delta_\ell~ n_\ell (kr)]Y_{\ell m}(\Omega)
\end{eqnarray}
where the parameter $k$ is defined from $E=\hbar^2k^2/(2\mu )$ and the functions $Y_{\ell m}(\Omega)$ represent the spherical harmonic functions defined at angles $\Omega$.
The spherical Bessel [$j_\ell(kr)$] and Von Neumann [$n_\ell(kr)$] functions correspond to the regular and irregular solutions of the Hamiltonian $\hat{H}_s$. 
The term $\delta_\ell$ denotes the phase shift induced by the potential $\hat{V}_s$.

In the following, the energy-normalized regular and irregular
solutions of the $H_c$ Hamiltonian are defined together with the
corresponding scattering matrices.
The symbol $\psi_c$ represents the regular functions at the origin and
$\chi_c$ represents the irregular functions.
Since the $V_c$ potential confines the motion in the
$x-$ and $y-$directions, then the parity in the $z$-direction, namely $\Pi_z$ is a good quantum number and hence the asymptotic form of the regular wave functions are expressed as even and odd solutions

\begin{equation}\label{eqPsi0eo}
\psi^{(e,o)}_{c,\alpha}(\mathbf{r})= \Phi_\alpha (x,y)\sqrt{\frac{\mu}{\pi\hbar^2k_\alpha }}\begin{cases} 
		 \cos (k_\alpha z),~~\rm{for}~~\Pi_z=+1\\
		 \sin (k_\alpha z),~~\rm{for}~~\Pi_z=-1,
		 \end{cases}
\end{equation}
and the irregular wave functions possess the following form in terms of even and odd solutions
\begin{equation}\label{eqChi0eo}
\chi^{(e,o)}_{c,\alpha}(\mathbf{r})= \Phi_\alpha (x,y)\sqrt{\frac{\mu}{\pi\hbar^2k_\alpha }}\begin{cases} 
		 \sin (k_\alpha |z|),~~\rm{for}~~\Pi_z=+1\\
		 -\frac{z}{|z|}\cos (k_\alpha |z|),~~\rm{for}~~\Pi_z=-1,
		 \end{cases}
\end{equation}
where the $\Phi_\alpha (x,y)$ are a complete basis of
orthonormal eigenfunctions of the transverse part of the $\hat{H}_c$ Hamiltonian and their specific form is dictated by the particular type of confining potential  $\hat{V}_c$.
The superscript notation $(e,o)$ in Eqs.~(\ref{eqPsi0eo}) and (\ref{eqChi0eo}) denotes the even ($\Pi_z=+1$) and
odd ($\Pi_z=-1$) solutions, respectively. The total energy is given by
the relation
$E=\hbar^2k_\alpha ^2/2\mu + E_\alpha $ where $E_\alpha $
is the energy of the transverse function $\Phi_\alpha (x,y)$.
These equations assume $E>E_\alpha $ assuring therefore that
the solutions possess an oscillatory behavior.
If the $E<E_\alpha $, then there are functions that are
real exponentials in $z$.

The scattering solutions of the full Hamiltonian $\hat{H}$ 
at energy $E$ can
be expressed at small distances as linear combinations 
of the eigenfunctions of
$\hat{H}_c$ Hamiltonian at this same energy $E$.
Moreover, if the scattering potential is an even function of
$z$, then the solutions separate into sums over the even
functions or sums over the odd functions.
In this situation, the asymptotic form of the exact wave
function of the Hamiltonian $\hat{H}$ can be written as
\begin{eqnarray}\label{eqKmat}
\psi^{(e,o)}_\alpha(\mathbf{r}) &=&\psi^{(e,o)}_{c,\alpha}(\mathbf{r})-\sum_\beta\chi^{(e,o)}_{c,\beta}(\mathbf{r})K^{(e,o)}_{\beta\alpha}
\end{eqnarray}
where the elements $K^{(e,o)}_{\beta\alpha}$ indicate
the $K$-matrix elements of even or odd states in the $z$-direction.
The $\alpha ,\beta$ indexes label the open channels
($E>E_{\alpha, \beta} $) meaning that the corresponding
wave function possess an oscillatory behavior as $|z|\to\infty$.
On the other hand asymptotically the closed channels
($E<E_\alpha $) are described via the following functions
\begin{equation}\label{eqUps}
\Upsilon^{(e)}_{\alpha}(\mathbf{r})=\Phi_{\alpha} (x,y)
\sqrt{\frac{\mu}{\pi\hbar^2\kappa_{\alpha} }}\exp (-\kappa_{\alpha} |z|)
\end{equation}
where the index $\alpha$ is here introduce in order
to separate the closed from open channels. The energy of
the closed channels is given by the relation
$E = -\hbar^2\kappa_{\alpha} ^2/2\mu + E_\alpha$.

Having specified the scattering eigenfunction of the
$\hat{H}_c$ Hamiltonian, the corresponding
$K$-matrix fulfills the following relation:
\begin{equation}
K_{\alpha\beta }= -\pi\braket{\psi_{c,\alpha} |\hat{V}_s|\psi_{\beta}}
\label{kmatnonvar}
\end{equation}
where $\alpha,~\beta$ label open channels.
In addition, the $\ket{\psi_c},~\ket{\psi}$ are the exact
solutions of $\hat{H}_c$ and $\hat{H}$ Hamiltonians, respectively.

Equation~(\ref{kmatnonvar}) can be expressed in a more general
way by substituting the exact relation
$\hat{V}_s\ket{\psi_\beta}=(E-\hat{H}_c)\ket{\psi_\beta}$
and integrating by parts twice.
Using this, the volume integral in
Eq.~(\ref{kmatnonvar}) can be recast as a surface integral that is a simpler 
expression for the corresponding $K$-matrix:
\begin{eqnarray}
K_{\alpha\beta}^{(e)}&=&-\frac{\hbar^2\pi}{\mu}\int\int W_z[\psi^{(e)*}_{c,\alpha}(\mathbf{r}),\psi^{(e)}_\beta(\mathbf{r})]dxdy,
\end{eqnarray}
where the term $W_z[\cdot]$ indicates the Wronskian with
respect to the $z$-direction and is evaluated at large enough
$|z|$ that the closed functions of Eq.~(\ref{eqUps}) are
effectively 0.
The closed channels are in the full wave function and could
lead to resonances but do not contribute to the surface
integrals at large $z$.

\subsection{Free particle and confining Green's functions}\label{secGF}

The Hamiltonians introduced in Eqs.~(\ref{eq2})
and (\ref{eq3}) permit us to straightforwardly define the
two corresponding Green's functions.
\begin{equation}\label{eqGdef}
\hat{G}_f\equiv\frac{1}{E-\hat{H}_f}
\qquad \hat{G}_c\equiv\frac{1}{E-\hat{H}_c},
\end{equation}
where $\hat{G}_c$ only includes the confining potential, and
$\hat{G}_f$ is the free particle Green's function
for the Hamiltonian with no potential energy.
The specific asymptotic boundary conditions in
Eq.~(\ref{eqGdef}) are determined by how the pole is handled.

The outgoing/incoming free particle Green's function
$\hat{G}_f$ in the position representation can be written as
\begin{equation}\label{eqGf}
\braket{\mathbf{r}_1|\hat{G}^\pm_f|\mathbf{r}_2}=G_f^\pm (\mathbf{r}_1,\mathbf{r}_2)=-\frac{\mu}{2\pi\hbar^2}
\frac{e^{\pm ikr_{12}}}{r_{12}},
\end{equation}
where $r_{12}=|\mathbf{r}_1-\mathbf{r}_2|$ and the ``$+$" (``$-$")
denotes the outgoing (incoming) Green's function.
The standing wave $G_f$ is derived simply by taking the real part
of either $G_f^\pm$.

The outgoing/incoming Green's function when only the confining
potential is present reads
\begin{eqnarray}\label{eqG0}
\braket{\mathbf{r}_1|\hat{G}^\pm_c|\mathbf{r}_2}&=&G_c^\pm (\mathbf{r}_1,\mathbf{r}_2) \cr
&=&\sum_{\alpha =1}^{\alpha_0}\frac{\mp i \mu}{\hbar^2 k_\alpha }
\Phi^*_\alpha (\boldsymbol{\rho}_1)\Phi_\alpha
(\boldsymbol{\rho}_2)e^{\pm ik_\alpha |z_1-z_2|}\cr 
&+& \sum_{\alpha_0+1}^{\infty}\frac{\mu}{\hbar^2 \kappa_\alpha }
\Phi_\alpha (\boldsymbol{\rho}_1)\Phi^*_\alpha
(\boldsymbol{\rho}_2)e^{-\kappa_\alpha |z_1-z_2|},
\end{eqnarray}
where $\boldsymbol{\rho}$ a vector lying into $x-y$ plane.
The index $\alpha_0$ denotes the last open channel for a given total energy $E$.
As with the $\hat{G}^\pm_f$, the standing wave $\hat{G}_c$
can be obtained from the real part of $\hat{G}_c^\pm$.
Note that this property does not hold in the case of a charged particle motion in magnetic fields.

One important observation is that every Green's function
diverges as $r_{12}\to 0$ in the same way for {\it any energy
not at a threshold of $H_c$}:
\begin{equation}
G_{any}(\mathbf{r}_1,\mathbf{r}_2) = -\frac{\mu}{2\pi\hbar^2}\frac{1}{r_{12}} + O(r_{12}^0)
\end{equation}
for small $r_{12}$; this result also does not depend on
whether the $G$ obeys standing wave or incoming/outgoing
boundary conditions.
This means the difference of any two Green's functions
is finite everywhere except for energies exactly at threshold.
In the expressions below, we will arrange to only have integrals
that involve the differences of Green's functions times functions
that are finite and non-zero in a finite region. Thus, our
expressions will automatically be finite without the need
for regularization procedures.

In this work, the standing wave solutions are
employed; therefore the corresponding Green's functions,
i.e. $\hat{G}_f$ and $\hat{G}_c$ denote the
principal value ones.

\section{Schwinger Variational Principle}\label{secSVP}

This section presents a derivation of the $K$-matrix
in terms of the Schwinger variational principle.
The main focus of this derivation is the standing wave
solution and the corresponding Green's function.
However, note that a similar derivation can be carried out
in cases of incoming/outgoing boundary conditions.

Using the Lippmann-Schwinger equation,
$\ket{\psi_\alpha} =\ket{\psi_{c,\alpha}} + \hat{G}_c \hat{V}_s\ket{\psi_{\alpha}}$,
and its hermitian conjugate in Eq.~(\ref{kmatnonvar}),
the following identity for the $K$-matrices is fulfilled
\begin{equation}\label{eqKdef}
K_{\alpha\beta}=-\pi \braket{\psi_{c,\alpha} |\hat{V}_s|\psi_{\beta}}
=-\pi\braket{\psi_\alpha  |\hat{V}_s|\psi_{c,\beta}},
\end{equation}
where the relation $\hat{G}_c^{\dagger}=\hat{G}_c$ is used.

A Schwinger-type variational expression for the $K$-matrix
can be obtained with the help of Eq.~(\ref{eqKdef}) and
the trial functions
$\ket{\psi^{(\rm{t})}}=\ket{\psi} +\ket{\delta\psi} $
of the full Hamiltonian $H$.
\begin{eqnarray}\label{eqSVP}
K^{\rm{var}}_{\alpha\beta}&\equiv & -\pi[
\braket{\psi_{c,\alpha}|\hat{V}_s|\psi^{(\rm{t})}_{\beta}}
+\braket{\psi^{(\rm{t})}_{\alpha}|\hat{V}_s|\psi_{c,\beta}}-\braket{\psi^{(\rm{t})}_{\alpha}|\hat{V}_s|\psi^{(\rm{t})}_{\beta}}\cr
&\null& +\braket{\psi^{(\rm{t})}_{\alpha}|\hat{V}_s~\hat{G}_c~\hat{V}_s|\psi^{(\rm{t})}_{\beta}}],
\end{eqnarray}
where the variational expression for the $K$-matrix,
$K^{\rm{var}}$, equals the exact $K$-matrix, i.e. Eq.~(\ref{kmatnonvar}),
plus terms of order $\delta\psi^2$:
\begin{eqnarray}
 K^{\rm{var}}_{\alpha\beta}&=&-\pi [\braket{\psi_{c,\alpha}|\hat{V}_s|\psi_\beta}-\braket{\delta\psi_\alpha|\hat{V}_s-\hat{V}_s\hat{G}_c\hat{V}_s|\delta\psi_\beta}]\cr
 &=& K_{\alpha\beta}+O(\delta \psi^2)
\end{eqnarray}
The trial functions
$\ket{\psi^{(\rm{t})}_\beta}$ are often written as linear
combinations of a basis set of functions $\ket{y_j}$.
\begin{equation}\label{trialexpan}
 \ket{\psi^{(\rm{t})}_\beta} = \sum_j \ket{y_j}C_{j\beta }.
\end{equation}

Substituting Eq.~(\ref{trialexpan}) in Eq.~(\ref{eqSVP}), the
coefficients $C_{j\alpha }$ are specified by requiring Eq.~(\ref{eqSVP})
to be variationally stable, i.e. $\partial K_{\alpha \beta}^{\rm{var}}/\partial C_{j\beta }=0$.
This yields the basis expansion version of the Schwinger
variational expression, $K_{\alpha \beta}^{\rm{var}}$, which reads
\begin{equation}\label{eqSVPb}
K^{\rm{var}}_{\alpha\beta}=-\pi\sum_{jj'}\braket{\psi_{c,\alpha}|\hat{V}_s|y_j}
[M^{-1} ]_{jj'}\braket{y_{j'}|\hat{V}_s|\psi_{c,\beta}}
\end{equation}
where
$M_{j'j}=\braket{y_{j'}|\hat{V}_s-\hat{V}_s\hat{G}_c\hat{V}_s|y_j}$.
In addition, the trial functions $\ket{\psi^{(\rm{t})}_\alpha}$
can be evaluated via the coefficients
$C_{\alpha ,j}$ which fulfill the following relation
\begin{equation}\label{eqSVPC}
C_{j\beta } = \sum_{j'}[M^{-1} ]_{jj'}\braket{y_{j'}|\hat{V}_s|\psi_{c,\beta }}.
\end{equation}

One of the main points of a variational
principle is that an exact $K$-matrix is obtained if
the $\ket{\psi^{(\rm{t})}}$'s equal the exact $\ket{\psi}$'s.
For the Schwinger variational principle, all the corresponding expressions
for the $K$-matrix involve terms of
$\hat{V}_s\ket{\psi^{(\rm{t})}}$.
This implies that the exact result for the $K$-matrix is
obtained for the Schwinger variational principle
even if the $\ket{\psi^{(\rm{t})}}$ satisfies
the somewhat looser condition
$V_s(\mathbf{r})\delta\psi (\mathbf{r})=0$ which will be
exploited in the next section.

In particular this condition can be satisfied at large and small distances $r$.
At large distances, the potential $\hat{V}_s$ vanishes
implying that the $\delta \psi(\mathbf{r})$ parts of the trial functions $\ket{\psi^{(\rm{t})}}$
which are nonzero in this region do not contribute to
the $K$-matrix calculation.
On the other hand, the trial functions at small distances can be chosen such that they are exact solutions of the full Hamiltonian $\hat{H}$ yielding therefore $\delta \psi(\mathbf{r})=0$ at small $r$.
This imposes a constraint on the choice of the trial
functions or the basis expansions that is employed in the next section.
With these choices, the exact $K$-matrix is obtained even if
the $\ket{\psi^{(\rm{t})}}$ is not an eigenfunction of the full
Hamiltonian $\hat{H}$ over all space.

An additional feature is that if $V_s(\mathbf{r}) \neq 0$
in a small region of space (and for a limited range of low energies), only the low-$\ell$ angular
momentum partial waves of  $\ket{\psi^{(\rm{t})}}$ are
needed to calculate the $K$-matrix.
This holds only away from high-$\ell$ resonances due to
the fact that as the angular momentum $\ell$ increases the
amplitude of the corresponding wave functions in the
region of $V_s(\mathbf{r}) \neq 0$ vanishes.
Therefore, these  high-$\ell$ states will not contribute
to the scattering information.

\section{K-matrix for two potentials: application of the Schwinger variational principle}\label{secII}

An important special case is when the potential
$\hat{V}_s=\hat{H}-\hat{H}_c$ is nonzero over a range
small enough that the potentials in $\hat{H}_c$ Hamiltonian
are considered effectively zero (or constant) over the
range of $\hat{V}_s$.
This will allow us to simplify all of the integrals in
Eqs.~(\ref{eqSVPb}) and (\ref{eqSVPC}) and in some cases
the corresponding results are simple analytic functions.
To reduce the amount of new material, we have restricted
the scattering systems to have no $z$-dependence to the
potential {\it and} for the confining potential to have
symmetry in $x,y$. Finally, an important consideration
for this section is to obtain expressions for the $M$-matrix
elements, Eq.~(\ref{eqSVPb}), as a simple analytic term
and a term involving an integral over finite functions.

\subsection{General expression of the K-matrix for
short-ranged spherical symmetric potential $\hat{V}_s$}

As discussed at the end of Sec.~\ref{secSVP}, only
low-$\ell$ angular momentum partial waves contribute to
the calculation of the $K$-matrix when $\hat{V}_s$ is assumed
to be short ranged.
Therefore, for such a short range potential only the
first two angular momentum states will be considered,
i.e. $\ell=0$ and $1$.
This approximation will allow us to obtain closed form
expressions of all the $K$-matrix elements in terms of
the $\ell =0$ and $1$ phase shifts.

Since the full potential $\hat{V}$ is symmetric in $z$, the
$K$-matrix separates into even- and odd-channel blocks.
For each parity state of the $\hat{H}_c$ Hamiltonian, only
one basis function is used in Eq.~(\ref{eqSVPb}) which is
the solution of $\hat{H}$ Hamiltonian.
This will provide the exact scattering information given that
the length scale of the potential terms $\hat{V}_s$ and
$\hat{V}_c$ in the full Hamiltonian $\hat{H}$ are well separated.
In other words, the length scale separation implies that
in the region of $\hat{V}_s \neq 0$ the states $\ket{\psi_s}$
(see Eq.~(\ref{eq3})) will not be modified by the confining
potential $\hat{V}_c$.

To be precise, if there is a potential that only scatters
$s$- and $p$-waves at some specified (low) energy, then the potential only extends over
a very short range.
The wave function in the region near the scatterer, at
position $\mathbf{r}_s$, is most easily represented in
spherical coordinates.
The difficulty will be to evaluate the integrals involving
the $\ket{\psi_c}$ and the corresponding $\hat{G}_c$ which
will often be most easily represented in a different
coordinate system.
The reason that the exact scattering information is obtained
from the Schwinger variational principle is that the
{\it exact} wave function of the full Hamiltonian (including
large $r$ boundary conditions) must have the form
$\psi (\mathbf{r}) = C \psi_s(\mathbf{r}) +\xi (\mathbf{r})$.
Note that $\psi_s$ is the spherical wave centered at $r_s$
that does not obey the large-$r$ boundary conditions of
the full Hamiltonian.
The function $\xi =\psi -C\psi_s$ only has $\ell\geq 2$
partial waves at $\mathbf{r}_s$ which implies
$V_s(\mathbf{r})\xi (\mathbf{r})= 0$ everywhere.
This means the exact $K$-matrix can be written as
\begin{equation}\label{eqSVPs}
K_{\alpha\beta}=-\pi \frac{\braket{\psi_{c,\alpha}|\hat{V}_s|\psi_s}\braket{\psi_s|\hat{V}_s|\psi_{c,\beta}} }
{\braket{\psi_s|\hat{V}_s-\hat{V}_s\hat{G}_c\hat{V}_s|\psi_s}},
\end{equation}
where each term is now a single matrix element.
The terms in the numerator give a finite result because the
$\hat{V}_s\ket{\psi_s}$ is non-zero only over a finite region.
Although the term in the denominator is not obviously
finite because the $\hat{G}_c$ diverges in proportion to
$1/|\mathbf{r}_1-\mathbf{r}_2|$, the derivation below shows that
this term can be written as convergent sums.

The evaluation of the term in the denominator of Eq.~(\ref{eqSVPs})
is complicated by the fact that the $\hat{G}_c$ is simple in a
different coordinate system from that of $\ket{\psi_s}$ and
$\hat{V}_s$ \cite{stephens1992jcp, lucchese1979jpb, lucchese1983pra, lino2000bjp, lucchese1983amopp}.
However, these complications can be avoided by adding and
subtracting the Green's function for a free particle, $\hat{G}_f$,
\begin{equation}\label{eqDen1}
\braket{\psi_s|\hat{V}_s-\hat{V}_s\hat{G}_c\hat{V}_s|\psi_s}= I_s + D_s
\end{equation}
with the new parameters $I_s$ and $D_s$ defined as
\begin{equation}\label{eqDen2}
I_s = \braket{\psi_s|\hat{V}_s-\hat{V}_s\hat{G}_f\hat{V}_s|\psi_s} = -\frac{1}{\pi}\tan\delta_\ell
\end{equation}
where the $\ket{\psi_s}$ indicates a state of a particular
angular momentum $\ell$ possessing the form of Eq.~(\ref{eqPsil}) and
\begin{equation}\label{eqDelG}
D_s=\braket{ \psi_s|\hat{V}_s\Delta \hat{G} \hat{V}_s|\psi_s} = \braket{\psi_s|\hat{V}_s (\hat{G}_f-\hat{G}_c)\hat{V}_s|\psi_s}
\end{equation}
These manipulations have the advantage that the $I_s$ is simply
related to the no confinement phase shift {\it and} the
$G_f(\mathbf{r}_1,\mathbf{r}_2)-G_c(\mathbf{r}_1,\mathbf{r}_2)$
function is {\it finite for all values} of $\mathbf{r}_1$ and $\mathbf{r}_2$.
Thus, the expression in Eq.~(\ref{eqDelG}) is finite because the
$\hat{V}_s\ket{\psi_s}$ is finite with a finite range and its square
is multiplying a finite function.
Note that a similar difference of Green's functions appears also in the methods presented in Refs.\cite{SlonimGreene1991,golubkovjphysb1984} yielding thus finite results. 

One important observation is that the $K$-matrix in Eq.~(\ref{eqSVPs})
is essentially separated in factors of matrix elements.
This is crucial since it allows us to treat each factor
separately in a pedagogical manner.
The following subsections are devoted to providing their
explicit analytical expressions.

\subsection{Matrix elements with $\hat{V}_s\ket{\psi_s}$}\label{subsec:secMEVP}

It is evident from Eq.~(\ref{eqSVPs}) that most of the integrals
involve terms of $\hat{V}_s \ket{\psi_s}$; thus this subsection
mainly focus on the integrals of the form
\begin{equation}\label{eqPsisC}
\braket{ \psi_{c}|\hat{V}_s|\psi_s}=\int_\tau \psi_{c}(\mathbf{r})V_s(\mathbf{r})\psi_s(\mathbf{r})d^3\mathbf{r},
\end{equation}
where  $\tau$ is the smallest volume that contains the region
where $V_s(\mathbf{r})\neq 0$.
Recall that the function $\psi_{c}(\mathbf{r})$ is an
eigensolution of the $\hat{H}_c$ Hamiltonian.
Since the $V_s(\mathbf{r})$ is nonzero only over a small region,
the integral should only be performed over a volume $\tau$ that
contains this region.

The Appendix \ref{secMEVPapp} gives the prescription
for evaluating these matrix elements for $\ell\leq 1$. The basic
idea is to convert the volume integral into a surface integral which
is possible due to the length scale separation between the range
of $\hat{V}_s$ and when the $\hat{V}_c$ becomes non-negligible. The
main results are:
\begin{equation}\label{eqNums}
\braket{\psi^{(e)}_{c,\alpha}|\hat{V}_s|\psi_{s,00}} =
\Phi^*_\alpha (0,0)\sqrt{\frac{2\pi}{k k_\alpha}}\left( -\frac{1}{\pi}\tan\delta_0\right)
\end{equation}
\begin{equation}\label{eqNump}
\braket{\psi^{(o)}_{c,\alpha}|\hat{V}_s|\psi_{s,10}} =
\Phi^*_\alpha (0,0)\sqrt{\frac{6\pi  k_\alpha}{k^3}}\left( -\frac{1}{\pi}\tan\delta_1\right)
\end{equation}
\begin{equation}\label{eqNump1}
\braket{\psi^{(e)}_{c,\alpha}|\hat{V}_s|\psi_{s,1\pm1}} =
-(\partial_\pm \Phi_\alpha )^*
 \Bigg|_0\sqrt{\frac{12\pi}{ k_\alpha k^3}}\left( -\frac{1}{\pi}\tan\delta_1\right)
\end{equation}
where $(\partial_\pm \Phi_\alpha )\equiv(\partial \Phi_\alpha (\mathbf{r})/\partial x\mp i\partial \Phi_\alpha (\mathbf{r})/\partial y )/2$.

\subsection{Analytic expressions for the $D_s$ matrix elements}\label{subsec:dselemen}

This subsection focuses on the $D_s$ matrix elements.
One of the main assumptions here is that the $\hat{V}_s$ potential is
centered symmetrically with respect to the confining potential; for
example, $\hat{V}_s$ is at the origin {\it and} $V_c(x,y)=V_c(y,x)$
and $V_c(x,y)=V_c(-x,y)$ etc.
This allows us to resolve possible challenges with respect to the
scattering of $\ell =1,|m|=1$.

The $D_s$ matrix elements for $\ell =0, m=0$ and for $\ell = 1,~m$
can be written in the compact form
\begin{equation}\label{dslm}
D_{s,\ell m }=\left(\frac{1}{\pi}\tan\delta_\ell \right)
\frac{\tan\delta_\ell}{[a_\perp k]^{2\ell +1}} \eta_{\ell m} (E)
\end{equation}
where the $a_\perp$ is a relevant length scale for the confinement
potential $\hat{V}_c$ and $\eta_{ \ell m}(E) $-function is defined
in terms of the $\Delta G$ defined in Appendix~\ref{secDsapp}.
For $\ell =0,~m=0$ and $\ell =1,~m$, they are defined as
\begin{eqnarray}\label{eqEtaGen}
\eta_{00}(E)&=&\frac{2\pi\hbar^2}{\mu}a_\perp \Delta G(0)\cr
\eta_{10}(E)&=&-\frac{6\pi\hbar^2}{\mu}a_\perp^3\Delta G_{zz}(0)\cr
\eta_{1\pm1}(E)&=&-\frac{6\pi\hbar^2}{\mu}a_\perp^3\Delta G_{\rho\rho}(0)
\end{eqnarray}

It is important to note that the expression for $\Delta G $
converges for non-zero $\Delta z$ without the need for regularization.
In order to investigate in detail the $\Delta z$ dependence at $\Delta \tilde{\rho}=\frac{\Delta \rho}{a_\perp} \ll 1$,
the function
$\Gamma (\boldsymbol{r}_1,\boldsymbol{r}_2,E)\equiv 2\pi\hbar^2a_\perp \Delta G(\boldsymbol{r}_{1} ,\boldsymbol{r}_2)/\mu$
is defined which reads
\begin{eqnarray}\label{eqGamma}
\Gamma (\boldsymbol{r}_1,\boldsymbol{r}_2,E)=
&-&2\pi\sum_{\alpha = 1}^{\alpha_o}a_\perp^2\Phi_\alpha (\boldsymbol{\rho}_1)\Phi_\alpha^* (\boldsymbol{\rho}_2)
\frac{\sin (\tilde{k}_\alpha |\Delta\tilde{z}|)}{\tilde{k}_\alpha}\cr
&+&2\pi\sum_{\alpha = \alpha_o+1}^{\infty}a_\perp^2\Phi_\alpha (\boldsymbol{\rho}_1)\Phi_\alpha^* (\boldsymbol{\rho}_2)
\frac{e^{-\tilde{\kappa}_\alpha |\Delta\tilde{z}|}}{\tilde{\kappa}_\alpha}\cr
&-&\frac{\cos (\tilde{k}\tilde{r}_{12})}{\tilde{r}_{12}},
\end{eqnarray}
where the $E_{\alpha_o}<E<E_{\alpha_o+1}$ defines the open and closed channels.
Parameters with a tilde have been scaled by the length scale
$a_\perp$ to become dimensionless:
$\Delta\tilde{z}=\Delta z/a_\perp$, $\tilde{k}=ka_\perp$,
$\tilde{k}_\alpha = k_\alpha a_\perp$, and
$\tilde{\kappa}_\alpha = \kappa_\alpha a_\perp$.
In the expression for $\Gamma$, the first term is the contribution
from the open channels to $\hat{G}_c$, the second term is the
contribution from the closed channels to $\hat{G}_c$, and the
last term is the contribution from $\hat{G}_f$.

The three $\eta_{\ell m}(E)$ (see Eq.~(\ref{eqEtaGen})) functions
are derived from the $\Gamma (\boldsymbol{r}_1,\boldsymbol{r}_2,E)$ in the limit $\Delta \tilde{z}\to 0^+$ and $\Delta \tilde{\rho} \ll 1$.
Physical considerations from Eq.~(\ref{eqdelGdelz}) means the
expansion of the $\Gamma (\tilde{\mathbf{r}}_1,\tilde{\mathbf{r}}_2,E)$
will have the form
\begin{eqnarray}\label{eqGammaapp}
\Gamma (\boldsymbol{r}_1,\boldsymbol{r}_2,E)&\simeq&\eta_{00}(E) + \frac{\eta_{10}(E)}{3}\tilde{z}_1\tilde{z}_2\cr
&+&\frac{\eta_{11}(E)+\eta_{1-1}(E)}{6}\tilde{\boldsymbol{\rho}}_1\cdot\tilde{\boldsymbol{\rho}}_2.
\end{eqnarray}

The physical meaning of the $\eta_{\ell m}(E)$ terms is that
they encapsulate the impact of the confining geometry,
i.e. $\hat{V}_c$, on the short-range spherical symmetric
potential $\hat{V}_s$.
Indeed, $\eta_{\ell m}$ conveys the information of each spherical
wave of $\ell m$ character induced by the $\hat{V}_s$ at short
distances is distributed over the asymptotic $\alpha$-channels
imposed by $\hat{V}_c$ at long distances.
Evidently, the explicit form of the $\eta_{\ell m}$ terms
depends on the specific type of the confining geometry.
Therefore, in the Appendix~\ref{secEtaapp}, the $\eta_{\ell m}(E)$
terms are evaluated in detail for two types of confining potentials, e.g.
the harmonic and the square well potential.

\begin{figure}
\includegraphics[scale=0.42]{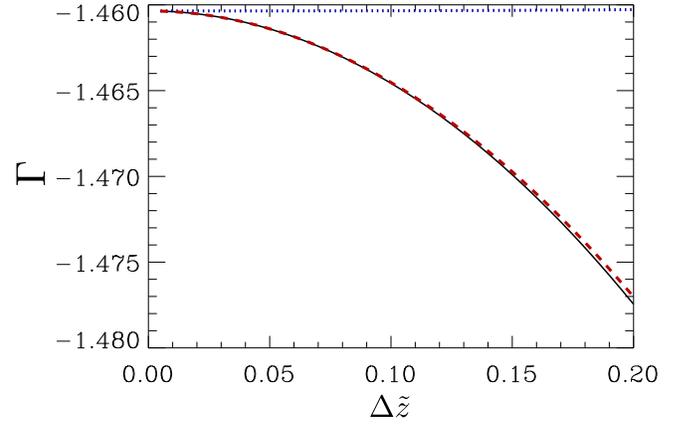}
\caption{(color online) The quantity $\Gamma$ of Eq.~(\ref{eqGammaapp}) as a function of
$\Delta z/a_\perp$for the confining potential of the harmonic
oscillator infinitesimally above the threshold energy $E=\hbar\omega_\perp$ where
$\boldsymbol{\rho}_1=\boldsymbol{\rho}_2=0$.
The black solid line is denotes the sum in Eq.~(\ref{eqGamHO}). The
red dashed line depicts the approximation Eq.~(\ref{eqGammaapp})
using Eq.~(\ref{eqEtaHO}). The blue dotted line indicates a numerical
extrapolation of
$\Gamma(\boldsymbol{r}_1,\boldsymbol{r}_2,E)\equiv\Gamma (\Delta \tilde{z}\to0^+, \Delta \tilde{\rho}\to0,E )\simeq [4\Gamma (\Delta\tilde{z},0,E)-\Gamma (2\Delta\tilde{z},0,E)]/3$.}
\label{fig1}
\end{figure}

\subsection{Test of Taylor series expansion of $\Delta G$}\label{secTSE}

Figure \ref{fig1} explores the various approximations on the
quantity $\Gamma$ which were discussed in Eqs.~(\ref{eqGamma})
and (\ref{eqGamHO}) considering that the confining potential is a harmonic one.
The solid line in Fig.~\ref{fig1} depicts $\Gamma -$quantity
from Eq.~(\ref{eqGamma}) at energy infinitesimally above the
threshold $E=\hbar \omega_\perp$ as a
function of $\Delta \tilde{z}$ for $\boldsymbol{\rho}_1=\boldsymbol{\rho}_2=0$.
The latter means that the azimuthal quantum number $m$ is set
equal to zero.
The red dashed line in Fig.~\ref{fig1} corresponds to Eq.~(\ref{eqGamHO})
which is obtained by calculating the $\eta_{00}(E)$ and
$\eta_{10}(E)$ using the method described in the
appendix \ref{secEtaapp}.
The red dashed and the black solid line are in excellent agreement,
especially for small $\Delta \tilde{z}$.
This highlights the validity of the approximations that were
considered in Eq.~(\ref{eqGamHO}).
In addition, the blue dotted line refers to $\eta_{\ell=0,m=0}(E)$
parameter which is defined as a linear combination of the
$\Gamma(\boldsymbol{r}_1,\boldsymbol{r}_2,E)$ functions canceling in this manner
terms of the order $|\Delta\tilde{z}|^2$ according to the relation $\Gamma(\boldsymbol{r}_1,\boldsymbol{r}_2,E)\equiv\Gamma (\Delta \tilde{z}\to0^+, \Delta \tilde{\rho}\to0,E )\simeq [4\Gamma (\Delta\tilde{z},0,E)-\Gamma (2\Delta\tilde{z},0,E)]/3$. 
Therefore, in Fig.~\ref{fig1} the dotted, solid and dashed
lines converge to the same finite value of $\eta_{\ell=0,m=0}$
without exhibiting any divergences.
A similar analysis can be carried out for the
$\eta_{\ell=1,m=1}(E)$ and $\eta_{\ell=1, m=\pm1}(E)$, and it shows 
excellent convergence between Eqs.~(\ref{eqGamma}) and (\ref{eqGamHO}).

\begin{figure}
\includegraphics[scale=0.42]{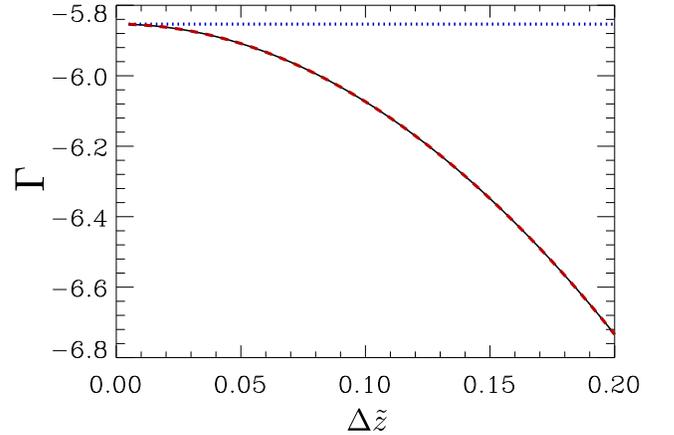}
\caption{(color online) The quantity $\Gamma$ of Eq.~(\ref{eqGammaapp}) for the
square well infinitesimally above the threshold energy $E=\hbar^2\pi^2/[\mu a_\perp^2 ]$
as function of $\Delta \tilde{z}$ for $\boldsymbol{\rho}_1=\boldsymbol{\rho}_2=0$.
The black solid line denotes the sum in Eq.~(\ref{eqGamSW}).
The red dashed line indicates the approximation Eq.~(\ref{eqGammaapp}) using Eq.~(\ref{eqEtaSW}).
The blue dotted line is a numerical extrapolation of $\Gamma(\boldsymbol{r}_1,\boldsymbol{r}_2,E)\equiv\Gamma (\Delta \tilde{z}\to0^+, \Delta \tilde{\rho}\to0,E )\simeq [4\Gamma (\Delta\tilde{z},0,E)-\Gamma (2\Delta\tilde{z},0,E)]/3$.}
\label{fig2}
\end{figure}

Similar to Fig.~\ref{fig1}, Fig.~\ref{fig2} investigates the
validity of  Eqs.~(\ref{eqGamma}) and (\ref{eqGamSW}) for the
case of a square well confining potential.
Fig.~\ref{fig2} mainly depicts the quantity $\Gamma$ as function
of $|\Delta\tilde{z}|$ for $\boldsymbol{\rho}_1=\boldsymbol{\rho}_2=0$
infinitesimally above the threshold energy $E=\hbar^2\pi^2/[\mu a_\perp^2 ]$.
The black solid line indicates Eq.~(\ref{eqGamma}) and the dashed line
corresponds to Eq.~(\ref{eqGamSW}).
It is apparent that both relations are in excellent agreement in
particular for $\Delta \tilde{z} \to 0$ suggesting in this manner
the regime of validity of the Eq.~(\ref{eqGamSW}).
In addition, note that the black solid and the red dashed line converge to
the same value as  $\Delta \tilde{z} \to 0$, i.e. $\eta_{00}(E)$.
Furthermore, the blue dotted line in Fig.~\ref{fig2} consists of a linear
combination of the $\Gamma(\boldsymbol{r}_1,\boldsymbol{r}_2,E)$ quantity such that terms of order
$|\Delta\tilde{z}|^2$ defined by the  relation $\Gamma(\boldsymbol{r}_1,\boldsymbol{r}_2,E)\equiv\Gamma (\Delta \tilde{z}\to0^+, \Delta \tilde{\rho}\to0,E )\simeq [4\Gamma (\Delta\tilde{z},0,E)-\Gamma (2\Delta\tilde{z},0,E)]/3$.
Evidently, in Fig.\ref{fig2} this linear combination (blue dotted line) is constant in the interval
of small $\Delta \tilde{z}$ and corresponds in essence to the value of $\eta_{00}(E)$.

\section{Scattering observables in terms of the Schwinger variational $K$-matrix}

This section utilizes key formulas developed in Section \ref{secII}
in order to explicitly connect the Schwinger variational $K$-matrix
of a two potential Hamiltonian with all the relevant scattering
observables such as the photoabsorption cross-section.
In addition, the transmission and reflection coefficients for
quasi-one dimensional Hamiltonians are derived in terms of the
Schwinger variational $K$-matrix.
This derivations permits us also to discuss the connection of the
present study with the theoretical framework of the local frame
transformation (LFT) theory. We reiterate that the derivation is
for the case of the confining potential being independent of $z$
and symmetric in $xy$; examples that do not follow these restrictions
will be presented elsewhere.

\subsection{The explicit analytical expressions of the $K$-matrix and its connection with the LFT theory}

In this subsection the formulas derived in the
subsections \ref{subsec:secMEVP} and \ref{subsec:dselemen} are
the main constituents of the Schwinger variational $K$-matrix
in Eq.~(\ref{eqSVPs}).
Recall that the short-range scatterer
is assumed to be placed at a symmetry point of the confining potential.
This implies that the $K$-matrix will be non-zero in blocks
depending on whether the asymptotic channel $\alpha$ couple
to $\ell m $ states or not.
Note that each $\alpha$-channel is described either by even
or odd function in $z$-direction, thus these states couple
to $\ell m$ spherically symmetric states if $\ell +m$ is either
even or odd, respectively.

In this manner, substituting Eqs.~(\ref{eqDen2}), (\ref{eqNums})
and (\ref{dslm}) in Eq.~(\ref{eqSVPs}) the $K$-matrix is provided
for a given $\ell m$ pair of quantum numbers, where the type of
confining geometry is not specified.
The latter permits us to provide a generic form of scattering
matrix elements which read

\begin{equation}\label{allklm}
 K^{\rm{(e,o})}_{\alpha \beta}=-\mathcal{U}_{\alpha ,\ell m}\frac{
\tilde{a}_\ell^{2\ell+1}(E)}{1+\tilde{a}_\ell^{2\ell+1}(E)\eta_{\ell m} (E)}
[\mathcal{U}^\dagger]_{\ell m,\beta},
\end{equation}
where the superscripts $\rm{(e,o)}$ refer to even or odd states
in $z$-direction, respectively.
Note that the terms $\tilde{a}_\ell^{2\ell+1}(E)=a^{2\ell+1}(E)/a^{2\ell+1}_\perp$ where 
$a_\ell^{2\ell+1}(E)=-\tan\delta_\ell(E)/k^{2 \ell +1}$
correspond to {\it $\ell$-wave energy-dependent scattering length
($\ell =0$)or volume ($\ell =1$)}. 
It is important to note that for some long range fields, ie. the polarization potential, the scattering volume diverges in the limit of zero total energy. 
On the other hand, in the problems of the present study the additional confining potential sets a lower limit in the total energy other than zero avoiding in this manner the divergence of the scattering volume.
However, in the regime of extremely weak confinement the lower bound in the total energy tends to zero yielding eventually a divergent scattering volume which must be analyzed more carefully.
$\eta_{\ell m}(E)$ is the next energy dependent parameter.
For specific $\ell m$ state $\eta_{\ell m}(E)$ parameter intrinsically possesses threshold singularities which occur below ($\alpha$)-channel thresholds (see Figs.\ref{fig3} and \ref{fig4}). 
The matrix elements $\mathcal{U}_{\alpha, \ell m}$ or its conjugate
transpose $[\mathcal{U}^\dagger]_{\ell m,\alpha}$ capture the coupling of the
short- and long-range physics.
In other words, the matrix $\mathcal{U}$ contains the information
that relates the short-range quantum numbers $\ell$ and
$m$ with the $\alpha$ ones which are fully dictated from
the confining potential, i.e.. long range physics.
\begin{equation}\label{lft}
\mathcal{U}_{\alpha,\ell m}= \begin{cases}
		  \sqrt{\frac{2\pi a_\perp }{k_\alpha}}\Phi_\alpha (0,0) & \ell=m=0 \\
		  \sqrt{6\pi a_\perp^3 k_\alpha}\Phi_\alpha (0,0) & \ell=1,~m=0 \\
		  -\sqrt{\frac{12\pi a_\perp^3 }{k_\alpha}}(\partial_{\pm}\Phi_\alpha )\big|_0 & \ell=1,~m=\pm1
		 \end{cases}
\end{equation}
where the operators are defined below Eq.~(\ref{eqNump1}).

At this point it becomes evident that Eq.~(\ref{lft}) is in
essence the {\it local frame transformation} (LFT) which permits the
the projection of the $\alpha-$states on the $\ell, m$ spherically
symmetric states. 
Note that $\mathcal{U}$ in Eq.~(\ref{lft}) and the LFT transformation $U$ in Refs. \cite{Greene1987pra,Zhang2013pra} are connected according to the relation $ \mathcal{U}_{\alpha,\ell m}=U_{\alpha,\ell m}(k a_\perp)^{\ell + 1/2}$.
Therefore, specifying the type of the confining potential, i.e.
harmonic oscillator or square well, one obtains the local
frame transformation either from cylindrical to spherical
coordinates \cite{Greene1987pra} or from Cartesian to spherical
coordinates \cite{Zhang2013pra}.
In particular, in Ref.~\cite{Zhang2013pra}, the $\ell=m=0$
derived LFT is identical to Eq.~(\ref{lft}).
Similarly, the LFT of Ref.~\cite{Greene1987pra} provides the
same results as Eq.~(\ref{lft}).
However, for completeness reasons it should be noted that there is a mistake
in Eq.~(27) of Ref.~\cite{Greene1987pra}, because strictly following
the notation of Ref.~\cite{Greene1987pra} would yield the new formula, namely 
$U_{nl}(m)=U_{q_nl}^{\rm{B=0}}(m)(2a)^{1/2}[n!/(n+|m|)!]^{-1/2}[\frac{1}{2}(2n + m +|m|+1)]^{-|m|/2}$.

In addition, the $K$-matrix equation, namely Eq.~(\ref{allklm}),
is exactly the same as the corresponding {\it physical}
$K$-matrices of Refs.\cite{GrangerBlume2004prl,Giannakeas2012pra,Zhang2013pra}.
However, there is a conceptual difference between the framework
presented here and the framework in  Refs.\cite{GrangerBlume2004prl,Giannakeas2012pra,Zhang2013pra}.
This difference arises from the fact that the $K$-matrices
in Refs.\cite{GrangerBlume2004prl,Giannakeas2012pra,Zhang2013pra}
did not obey the physical boundary conditions for the closed
channels; therefore a closed channel elimination was employed
in order to obtain the physical $K$-matrix which obeys the
proper boundary conditions.
This procedure yield $\eta_{\ell m}(E)$ functions which
involve only the sum parts of Eqs.~(\ref{eqGamHO}) and
(\ref{eqGamSW}). Thus, the LFT leads to divergences.
This singular behavior was removed in a secondary step that required thoughtful implementation of
{\it auxiliary} techniques of regularization
, such as Riemann zeta function
regularization \cite{GrangerBlume2004prl, Giannakeas2012pra} or
residue regularization \cite{Zhang2013pra}.
On the other hand in the present theoretical framework no
such techniques need to be employed since by definition
$\eta_{\ell m}(E)$ functions are finite.
However, note that $\eta_{\ell m}(E)$ possess intrinsic threshold singularities at energies 
infinitesimally below the channel thresholds of the confining potential.
Furthermore, the Schwinger variational formalism provides
an intuitive understanding of the regularization techniques.
As is shown in the Appendix~\ref{secEtaapp},
the $\eta_{\ell m}(E)$ functions consist of a difference of a
sum and its integral representation, whereas the latter originates
from the free space Green's function.
This piece of information is absent from the physical
$K$-matrix in the LFT theory and it is incorporated by
various regularization schemes.
Perhaps a comparison between the derivation of the LFT and
the present results could lead to a deeper understanding of the
source of the divergences in the LFT, and deserves to be studied in the future. 

\subsection{Photoabsorption Cross-section}

Since the Schwinger variational $K$-matrix is defined one can
also define the electric dipole matrix elements and consequently the
photoabsorption cross section.
For example, these relations are needed for the treatment
of negative ion photodetachment in the present of external fields.
Initially, the expression for the trial function coefficients
on the basis of the $\ket{\psi_s}$, i.e.. Eq.~(\ref{eqSVPC}) are defined as
\begin{equation}\label{eqCalp}
C_{\ell m,\alpha} =\frac{\mathcal{U}^\dagger_{\ell m,\alpha}/[ka_\perp ]^{\ell +1/2}}
{1+\tilde{a}_\ell^{2\ell+1}(E)\eta_{\ell m}(E)}.
\end{equation}

Since we are interested in computing the cross
section that describes, for example, the excitation of a
negative ion by photon absorption, the dipole matrix elements
at small distances are defined by the relation
$D_{lm}(E)=\braket{\psi_{\ell m}|\hat{\varepsilon}\cdot \hat{r}|\psi_{\rm{init}}}$.
Note that the term $\hat{\varepsilon}\cdot \hat{r}$ is the dipole
operator with $\hat{\varepsilon}$ denoting the polarization vector
and the state $\psi_{\rm{init}}$ corresponds to the initial state
of the negative ion.
Having defined the dipole matrix elements at short distances,
one can derive the dipole matrix elements that describe the
transition from the initial state to the $\alpha-th$ state
which includes the confining potential.
Therefore, the dipole matrix elements read
\begin{equation}\label{eqDalph}
D_\alpha = D_{\ell m}(E)C_{\ell m,\alpha}=
\frac{\tilde{D}_{\ell m}(E)\mathcal{U}^\dagger_{\ell m,\alpha}}
{1+\tilde{a}_\ell^{2\ell+1}(E)\eta_{\ell m} (E)}
\end{equation}
where the element $\tilde{D}_{\ell m}(E)=D_{\ell m}(E)/(ka_\perp)^{\ell +1/2}$
typically has less energy dependence than $D_{\ell m}(E)$.

In order to obtain photoabsorption cross sections, we define
the dipole matrix elements which describe the transitions from
the initial state to the incoming wave final state that
possesses only outgoing waves in the $\alpha-th$ channel
yielding the following expression:

\begin{equation}
D^-_\alpha = \sum_\beta D_\beta [(1-iK)^{-1}]_{\beta \alpha },
\end{equation}
where the $K$-matrix has rank one for each $\ell m$ pair, with
the relevant eigenvector being proportional to the $D_\alpha$.
This allows a relatively simple expression for the dipole matrix elements.
\begin{equation}
D^-_\alpha = \mathcal{U}_{\alpha ,\ell m}\frac{\tilde{D}_{\ell m}(E) }{1+
\tilde{a}_\ell^{2\ell+1}(E)(\eta_{\ell m} (E)-iN_{\ell m}^2)},
\end{equation}
with
\begin{equation}\label{eqUN2}
N_{\ell m}^2=\sum_{\alpha =0}^{\alpha_o} |\mathcal{U}_{\alpha ,\ell m}|^2
\end{equation}
where the summation index $\alpha$ indicates a sum over only
open channels.
Then the total probability is proportional to $|D^-|^2=\sum_\alpha  |D^-_\alpha |$
which is the sum over the partial
absorption terms $P_\alpha= |D^-_\alpha |^2$. The sum reduces to the relation
\begin{equation}\label{eqPabsTot}
|D^-|^2 =\frac{N_{\ell m}^2\tilde{D}_{\ell m}^2(E)}{
[1+\tilde{a}_\ell^{2\ell+1}(E)\eta_{\ell m} (E)]^2
+[\tilde{a}_\ell^{2\ell+1}(E)N_{\ell m}^2]^2}.
\end{equation}

\subsection{Transmission and reflection coefficients}

This subsection contains expressions for
the transmission and reflection coefficients
of quasi-one dimensional Hamiltonians in terms of the
Schwinger variational $K$-matrix in Eq.~(\ref{allklm}).
This situation models the scattering aspects
of two-body collisions in the presence of a transverse 
harmonic potential or else for collisional complexes of light particles
being scattered by heavy ones in the presence of a confining
transverse potential.

By superposing the even and odd wave functions, the solution
with an incoming wave in the $\alpha-th$ channel can be obtained when
$s$- and/or $p$-wave scattering dominates.
\begin{eqnarray}
\psi^+_\alpha(\mathbf{r}) &=& \frac{i}{2}\sum_\beta \psi_\beta^{(o)}(\mathbf{r})[(1 - i K^{(o)})^{-1}]_{\beta\alpha} \cr
&+&\frac{1}{2}\sum_\beta \psi_\beta^{(e)}(\mathbf{r})[(1 - i K^{(e)})^{-1}]_{\beta\alpha},
\end{eqnarray}
where the $\psi^+_\alpha(\mathbf{r})$ only has a wave incoming 
from negative $z$ in channel $\alpha$.
In the case of identical particle collisions having only $s$-wave ($p$-wave) character, the odd (even) $K$-matrix will vanish, namely $K^{(o)}=0$ ($K^{(e)}=0$).
On the other hand in the case of distinguishable particles, $s$- and $p$-waves could potentially both be present, and then both even and odd parity $K$-matrices can contribute in the $\psi^+_\alpha(\mathbf{r})$ wave function.

From the preceding equation one can define the matrices of the corresponding reflection and
transmission amplitudes which fulfill the relations
\begin{equation}
r=\frac{iK}{1-iK}\qquad t=\frac{1}{1-iK},
\end{equation}

In the case of $s$-wave scattering the reflection and transmission amplitudes can be expressed in terms of the energy dependent scattering length and the matrix $\mathcal{U}$:
\begin{eqnarray}\label{eqRT}
r_{\beta\alpha} &=& -\mathcal{U}_{\alpha ,00}\frac{i \tilde{a}_0(E)}
{1+\tilde{a}_0(E)(\eta_{00} (E) + iN_{00}^2)} [\mathcal{U}^\dagger]_{00,\beta}\cr
t_{\beta\alpha}&=&\delta_{\beta\alpha} + r_{\beta\alpha}
\end{eqnarray}
Note that similar expressions can be obtained for $p$- and $s+p$-wave scattering but it is straightforward to derive them if needed.

From these matrix elements, the scattering probabilities can be obtained.
The probability for an incoming wave in channel $\alpha $ to
reflect into channel $\beta$ is $|r_{\alpha\beta}|^2$ while
the probability to transmit into channel $\beta$ is $|t_{\alpha\beta}|^2$.
Note that the reflection and transmission probabilities are the
same except when $\beta =\alpha$.
This is understandable on physical grounds: a zero range
potential scatters equally into $+z$ and $-z$ but the
$\beta =\alpha$ channel has different interference with the
incoming wave in the $+z$ and $-z$ directions.

\begin{figure}
\includegraphics[scale=0.38]{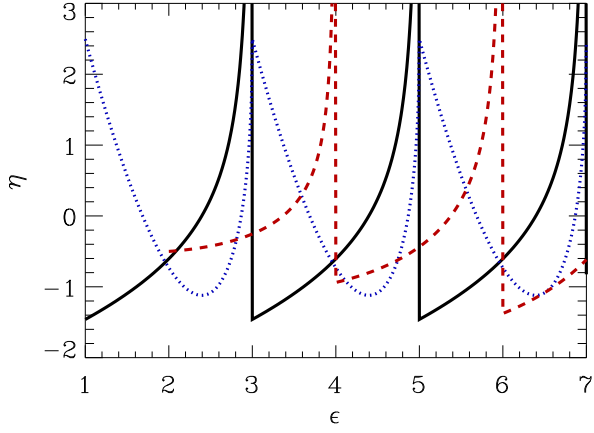}
\caption{The $\eta_{00}$ (black solid line), $\eta_{10}$ (blue dotted line)
and $\eta_{11}/20$ (red dashed line)
for harmonic oscillator confinement
as a function of the scaled energy $\epsilon = E/(\hbar\omega_\perp )$.
}
\label{fig3}
\end{figure}

\section{Results and discussion}

This section focuses on the various aspects and insights provided
by the present framework of the Schwinger variational $K$-matrix
approach and its comparison with the LFT results.
In addition, a set of selected applications of the current approach
is thoroughly discussed to highlight its robustness and clarity.
A first test bed is taken to be the study of photodetachment processes in
the presence of an external confining field where the closed channel
physics is taken into account, thus going beyond previous
studies \cite{Greene1987pra}.

Recall that the closed channel physics is fully encapsulated in
the $\eta_{\ell m}(E)$ functions which depend on the particular
$\ell m$ states of the short range potential and on the specific
type of the confining potential.

In other words, $\eta_{\ell m}(E)$ corresponds to a collective
parameter, which emerges from the
strong mutual coupling of all the closed channels caused by the
spherically symmetric short range potential at small distances.
In terms of the LFT theory, this collective parameter is described
by the eigenvalues of the closed-closed partition of the $K$-matrix
and, as was mentioned above, these $K$-matrices are of rank
1 despite the fact that their dimension is $M\times M$ with
$M\to \infty$. This particular property ensures that there exists
only one nonzero eigenvalue representing the effect of all of the closed channels.
Incorporating this concept in photodetachment processes yields
non-trivial resonant features; therefore, in the following
subsection the energy dependence and broad features of the
$\eta_{\ell m}(E)$ functions are studied for various confining potentials.

\subsection{Behavior of $\eta_{\ell m}(E)$}

\subsubsection{Harmonic and infinite square well confining potentials}

\begin{figure}[t!]
\includegraphics[scale=0.4]{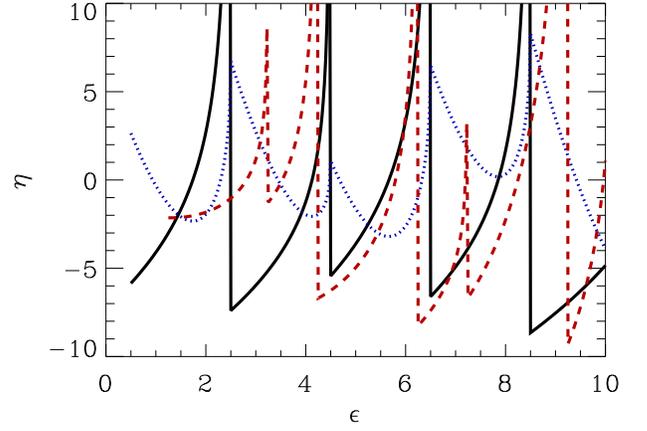}
\caption{The $\eta_{00}$ (black solid line), $\eta_{10}/50$ (blue dotted line)
and $\eta_{11}/300$ (red dashed line)
for infinite square well confinement
as a function of the scaled energy
$\epsilon = E/(4\hbar^2\pi^2/[2\mu a_\perp^2] )$.
}
\label{fig4}
\end{figure}

Figures \ref{fig3} and \ref{fig4} depict the quantity
$\eta_{\ell m}(E)$ as a function of the dimensionless
energy $\epsilon$ for the harmonic and infinite square well
confining potentials, respectively.
Note that the scaled energies $\epsilon$ fulfill the relations
$\epsilon=E/\hbar \omega_\perp$ and
$\epsilon=E/(4\hbar^2\pi^2/[2\mu a_\perp^2] )$
for the harmonic (Appendix~\ref{secGHO}) and square well
(Appendix~\ref{secGSW}) potentials, respectively.

The expressions for $\eta_{\ell m}$ for the harmonic and square well potential are given in Eqs.~(\ref{eqEtaHOapp}) and (\ref{eqEtaSWapp}), respectively. 
These parameters are expressed in terms of sums
and integrals involving the scaled momenta and $\Delta\tilde{z}$ which can be simplified to show only the dependences
on the quantum numbers.
For the harmonic oscillator, Eq.~(\ref{eqEtaHOapp})
becomes

\begin{eqnarray}\label{eqEtaHO}
\eta_{00}&=&\sum_{n>\nu_0}^\infty\frac{e^{-2\sqrt{n-\nu_0}\Delta\tilde{z}}}
{\sqrt{n-\nu_0}}-\int_{\nu_0}^\infty
\frac{e^{-2\sqrt{n-\nu_0}\Delta\tilde{z}}}
{\sqrt{n-\nu_0}} dn\cr
\eta_{10}&=&-12\sum_{n>\nu_0}^\infty\sqrt{n-\nu_0}\;e^{-2\sqrt{n-\nu_0}\Delta\tilde{z}}\cr
&\null&+12\int_{\nu_0}^\infty\sqrt{n-\nu_0}\;e^{-2\sqrt{n-\nu_0}\Delta\tilde{z}}dn
\cr
\eta_{1\pm 1}&=&\sum_{n>\nu_1}^\infty\frac{(n+1)
e^{-2\sqrt{n-\nu_1}\Delta\tilde{z}}}
{\sqrt{n-\nu_1}}\cr
&\null&-\int_{\nu_1}^\infty
\frac{(n+1)e^{-2\sqrt{n-\nu_1}\Delta\tilde{z}}}
{\sqrt{n-\nu_1}} dn
\end{eqnarray}
where the limit $\Delta \tilde{z} \to 0^+$ is understood in these expressions.
$nu_m$ is given by $\nu_m=(\epsilon-1-|m|)/2$ and it relates with momenta $\tilde{\kappa}_{n,m}$ in Eq.~(\ref{eqEtaHOapp}) according to the relation $\tilde{\kappa}_{n,m}=2\sqrt{n-\nu_m}$. 
For the square well confinement,
Eq.~(\ref{eqEtaSWapp}) becomes
\begin{eqnarray}\label{eqEtaSW}
\eta_{00}&=&4\sum_{\alpha_0^2 >\epsilon}^\infty
\frac{e^{-2\pi\sqrt{\alpha_0^2-\epsilon}\Delta\tilde{z}}}
{\sqrt{\alpha_0^2-\epsilon}}-4\int_{\sqrt{\epsilon}}^\infty
\frac{e^{-2\pi\sqrt{\alpha^2-\epsilon}\Delta\tilde{z}}}
{\sqrt{\alpha^2-\epsilon}}\frac{\pi}{2}\alpha d\alpha\cr
\eta_{10}&=&-48\pi^2\sum_{\alpha_0^2 >\epsilon}^\infty
\sqrt{\alpha_0^2-\epsilon}\;e^{-2\pi\sqrt{\alpha_0^2-\epsilon}\Delta\tilde{z}}
\cr
&\null&+48\pi^2\int_{\sqrt{\epsilon}}^\infty
\sqrt{\alpha^2-\epsilon}\;e^{-2\pi\sqrt{\alpha^2-\epsilon}\Delta\tilde{z}}
\frac{\pi}{2}\alpha d\alpha\cr
\eta_{1\pm 1}&=&48\pi^2\sum_{\alpha_1^2 >\epsilon}^\infty
\frac{(\alpha_x+1)^2e^{-2\pi\sqrt{\alpha_1^2-\epsilon}\Delta\tilde{z}}}
{\sqrt{\alpha_1^2-\epsilon}}\cr
&\null&-48\pi^2\int_{\sqrt{\epsilon}}^\infty
\frac{e^{-2\pi\sqrt{\alpha^2-\epsilon}\Delta\tilde{z}}}
{\sqrt{\alpha^2-\epsilon}}\frac{\pi}{4}\alpha^3 d\alpha
\end{eqnarray}
where where the limit $\Delta \tilde{z} \to 0^+$ is understood in these expressions. $\alpha_m$ is given by $\alpha_m^2=(\alpha_x+[|m|+1]/2)^2+(\alpha_y+1/2)^2$ with $\alpha_x,~~\alpha_y$ independently being $0,1,2,\ldots$.
$\alpha_m^2$ is related with the momenta $\tilde{\kappa}_{\alpha m}$ in Eq.~(\ref{eqEtaSWapp}) according to the expression $\tilde{\kappa}_{\alpha,m}=2 \pi \sqrt{\alpha_m^2-\epsilon}$.
In the integrals,
the double integrals $d\alpha_xd\alpha_y$ have been converted to polar
coordinates and the integral over angle from $0$ to $\pi /2$ has been
carried out.

Figure~\ref{fig3} shows the results of Eq.~(\ref{eqEtaHO}) for harmonic
oscillator confinement for $(\ell m)=[(00),(10),(11)]$ states.

We identify that the parameters $\eta_{00}(E)$ and $\eta_{10}(E)$ from Eq.~(\ref{eqEtaHO}) correspond to the Hurwitz Riemann functions $\zeta(1/2,3/2-E/2\hbar \omega_\perp)$ and $\zeta(-1/2,3/2-E/2\hbar \omega_\perp)$, respectively.
Note that exactly the same formulas was shown in Refs.\cite{olshanii1998prl,GrangerBlume2004prl}.
Furthermore, as was shown also in Ref. \cite{hess15arxiv},
the $\eta_{00}(E)$ (see black solid line in Fig.\ref{fig3}) and $\eta_{10}(E)$ (see blue dotted line in Fig.\ref{fig3})
are periodic functions of
$\epsilon = E/(\hbar\omega )$: $\eta_{\ell m} (\epsilon +2)=\eta_{\ell m} (\epsilon )$.

However, the $\eta_{11}(E)$ function (see red dashed line in Fig.\ref{fig3}) starts at a higher threshold
energy because the lowest transversal state for $m=1$ is an $\hbar\omega$ higher in energy than for $m=0$ and it is not periodic in $\epsilon$.
Actually, the $\eta_{11}(E)$ consists of a sum of the periodic
functions $\eta_{00}(E)$ and $\eta_{10}(E)$. 
However, the corresponding $\eta_{00}(E)$
term is proportional to the energy, which explains the non-periodic character of $\eta_{11}(E)$.

The $\eta_{00}$ and $\eta_{1\pm 1}$ functions diverge
as the energy approaches a channel from below due to threshold singularities whereas the $\eta_{10}$ remains
finite at each channel threshold. 
More specifically, the $\eta_{00}$ and $\eta_{1\pm 1}$
diverge due to the terms of
the $1/\kappa_\alpha$ in the summation whereas the $\eta_{10}$
does not exhibit such features at threshold because the summation
involves terms proportional to $\kappa_\alpha$.
Furthermore, at each threshold,
the value $\eta_{10}=2.494632...$ which is the same as in
Refs. \cite{GrangerBlume2004prl, hess2014pra}.
Similarly, the $\eta_{00}$ parameter acquires the value
$\eta_{00}=-1.460354...$ by approaching the thresholds from above
which is the same as in Refs. \cite{demkovjetp1966, olshanii1998prl}.
The $\eta_{11}$ function 
has the values $\eta_{11}=-10.00944$
 at the lowest threshold and $-18.77157$
at the first excited threshold when approaching those thresholds from above.
On the other hand, when the threshold is approached from below the
$\eta_{11}$ function diverges and this behavior arises from
the singular behavior of the $\eta_{00}$.

Figure \ref{fig4} illustrates  the $\eta_{\ell m}(E)$ functions for
the infinite square well potential where $(\ell m)=[(00),(10),(11)]$
states are considered.
More specifically, the functions $\eta_{00}$ (black solid line),
$\eta_{10}/50$ (blue dotted line) and $\eta_{11}/300$ (red dashed line)
are shown versus the scaled energy
$\epsilon = E/(4\hbar^2\pi^2/[2\mu a_\perp^2] )$.
In contrast to Fig.~\ref{fig3}, none of the $\eta$'s are periodic
functions of $\epsilon$.
As with Fig.~\ref{fig3}, the $\eta_{00}$ and $\eta_{11}$ go to
$\infty$ as the thresholds are approached from below whereas
they acquire a finite negative value as the thresholds are
approached from above.
This behavior arises from a similar analysis to that for
Fig.~\ref{fig3} but now involving the individual terms in Eq.~(\ref{eqEtaSW}).
As in the harmonic oscillator example, $\eta_{11}$
starts at a higher threshold energy because the transverse
functions must have a node to contribute, since the azimuthal
quantum number is $m=1$.
Our threshold value for $\eta_{00}=-5.853459...$ is in excellent
agreement with Ref. \citep{Zhang2013pra} result from the R-matrix
calculation but is $\sim 0.2$\% different from the result of
their regularized analytical summation.
In addition, from Fig.~\ref{fig4} the threshold values of the
functions $\eta_{10}$ and $\eta_{11}$ acquire the values
$\eta_{10}=132.098...$ and $\eta_{11}=-642.613...$, respectively.
Note that $\eta_{10}$ has been divided by a factor of
50 and $\eta_{11}$ by a factor of 300 to plot them on the same graph.
This indicates that the effect of the confinement on $p$-wave
scattering will typically be larger for the infinite square well
than for the harmonic oscillator.

\subsubsection{Infinite Rectangular Well Confinement}

Due to the simplicity of the current formalism one can explore
the properties of various confining geometries. In this section, the
case of a rectangular well confinement is considered.
This particular geometry permits us to study the impact of
lifting the degeneracy of the square-well potential.
Furthermore, in the following the discussion is restricted on $s$-wave scattering which couples only on even $z$-parity states of the rectangular well and/or on $p$-wave collisions which couples odd $z$-parity states of the rectangular well.

The rectangular potential well is defined as follows: $V(x,y)=0$
when $|x|<a_\perp /2$ {\it and} $|y|<a_\perp /(2c)$ and
$V(x,y)=\infty$ otherwise.
Note that the parameter $c$ controls the degree of
asymmetry in the potential.
The transverse eigenfunctions of the corresponding
confining Hamiltonian $\hat{H}_c$ are well known and close to
the origin obey the relation
\begin{equation}
\Phi_{\alpha_x,\alpha_y} (0,0) = \frac{2(-1)^{\alpha_x+\alpha_y}}{a_\perp\sqrt{c}}
\end{equation}
for all $(\alpha_x,\alpha_y)$ that are non-zero at the origin.
The corresponding eigenenergies for the states that are nonzero
at the origin can be written as
\begin{equation}
E_{n_xn_y}=\frac{\hbar^2\pi^2}{2\mu a_\perp^2}(n_x^2
+ [n_y/c]^2)
\end{equation}
where $n_x=2\alpha_x+1$ and $n_y=2\alpha_y+1$ where the
$\alpha_x,\alpha_y$ can independently be $0,1,2,3,...$.

\begin{figure}[t!]
\includegraphics[scale=0.45]{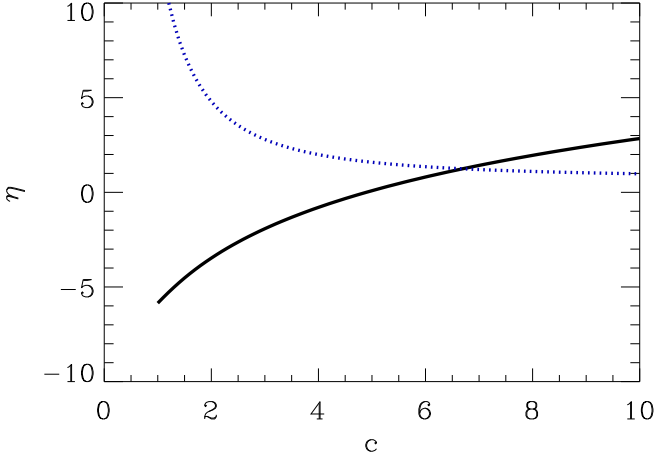}
\caption{(color online) The threshold value of $\eta_{00}$ (black solid line) and
$\eta_{10}/10$ (blue dotted line) for the infinite rectangular well
as a function of the ratio of confining length scales, $c$
defined as
$|x|<a_\perp /2$ {\it and} $|y|<a_\perp /(2c)$. Note $\eta_{00}$
goes from negative to positive at $c\simeq 4.89$.
}
\label{fig5}
\end{figure}

Since we are interested in $s-$ and $p$-wave scattering which respectively couple to even and odd-parity states of the rectangular well the prescription given in the appendix \ref{secGSW} is followed in order to obtain the corresponding parameters.
Namely, the $\eta_{00}$ and $\eta_{10}$ parameters fulfill the following relations:
\begin{eqnarray}\label{eqEtaRW}
\eta_{00}&=&\frac{4}{c}\sum_{\alpha_0^2 >\epsilon}^\infty
\frac{e^{-2\pi\sqrt{\alpha_0^2-\epsilon}\Delta\tilde{z}}}
{\sqrt{\alpha_0^2-\epsilon}}-4\int_{\sqrt{\epsilon}}^\infty
\frac{e^{-2\pi\sqrt{\alpha^2-\epsilon}\Delta\tilde{z}}}
{\sqrt{\alpha^2-\epsilon}}\frac{\pi}{2}\alpha d\alpha\cr
\eta_{10}&=&\frac{-48\pi^2}{c}\sum_{\alpha_0^2 >\epsilon}^\infty
\sqrt{\alpha_0^2-\epsilon}\;e^{-2\pi\sqrt{\alpha_0^2-\epsilon}\Delta\tilde{z}}
\cr
&\null&+48\pi^2\int_{\sqrt{\epsilon}}^\infty
\sqrt{\alpha^2-\epsilon}\;e^{-2\pi\sqrt{\alpha^2-\epsilon}\Delta\tilde{z}}
\frac{\pi}{2}\alpha d\alpha
\end{eqnarray}
where where the limit $\Delta \tilde{z} \to 0^+$ is understood in these expressions and $\alpha_0$ obeys the relation $\alpha_0^2=(\alpha_x+1/2)^2+(\alpha_y+1/2)^2$. In the integrals,
the double integrals $d\alpha_xd\alpha_y$ have been converted to polar
coordinates and the integral over angle from $0$ to $\pi /2$ has been
carried out.
Our numerical tests demonstrate that the resulting equations for
$\Gamma$ and $\eta_{00}$,$\eta_{10}$ are in agreement
and produce results similar to
those shown in Fig.~\ref{fig1} and \ref{fig2}.
The expressions in Eq.~(\ref{eqEtaRW}) use the fact that the
sums and integrals coincide in the continuum limit.
To see that this is the case, note that the sums in $\alpha_y$
have a factor of $c$ more states for each interval of energy
which cancel the factor of $1/c$ multiplying each sum.

\begin{figure}
\includegraphics[scale=0.4]{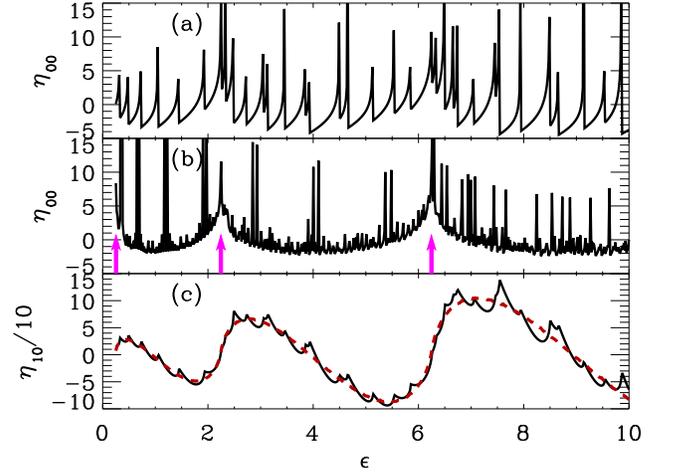}
\caption{(color online) The $\eta_{00}$ parameter is depicted in panels (a) and (b) for $c=5$ and $c=40$, respectively.
The $\eta_{10}/10$ parameter is depicted in panel (c) for $c=5$ (black solid line),
and $c=40$ (red dashed line). All the $eta$-parameters are plotted
as a function of the scaled energy
$\epsilon = E/(4\hbar^2\pi^2/[2\mu a_\perp^2] )$. The magenta arrows in panel (b) show the thresholds  
for the tight confinement in one direction, ie $c\gg1$, which acquire the values $\epsilon = 1/4$, $9/4$,
$25/4$, ...}
\label{fig6}
\end{figure}

Figure \ref{fig5} shows the $\eta_{00}$ (black solid line) and
$\eta_{10}/10$ (blue dotted line) from Eq.~(\ref{eqEtaRW}) as a
function of the ratio of $y$- to $x$-length scale at the
threshold energy $E=E_{1,1}$.
The $\eta_{10}$ stays positive for the values shown but the
$\eta_{00}$ changes sign for $c\simeq 4.89$.
At this value of $c$, the effect of the infinite closed
channels disappears for any finite value of the scattering length.
In other words, the scattering behaves
as if there were no confining potential at all; a similar effect
also exists in the harmonic oscillator confining potential which
occurs at specific values of energies away from the channel
thresholds \cite{hess2014pra}.
Over the range shown in Fig.\ref{fig6} at large values of $c$ the threshold value of $\eta_{00}$ appears to be
proportional to $\sqrt{c}$ while the $\eta_{10}$ appears to be
a constant plus a function proportional to $1/\sqrt{c}$.  

Figure \ref{fig6} depicts the energy dependence of $\eta_{00}$ and
$\eta_{10}/10$ parameters for different values of $c$.
As $c$ becomes larger, the corresponding Hamiltonian changes
from a quasi-one dimensional to a quasi-two dimensional one.
Since the $\eta_{00}$ diverges just below each threshold, these
curves exhibit many points of divergence corresponding to each threshold.
However, an overall pattern can be seen in panels (b) and (c).
This pattern is clearer in the plot of $\eta_{10}/10$ [see red dashed line in Fig.\ref{fig6}(c)] which does
not diverge at threshold.
In Figs.\ref{fig6} (a) and (b) the $c$ parameter acquires the values $c=5$ and $c=40$.
In these panels is observed that as $c$ increases the system behaves more like it is only
confined in the $x$-direction and therefore those thresholds 
[see the magenta arrows at $\epsilon = 1/4$, $9/4$, $25/4,\ldots$ if Fig.\ref{fig6}(b)]
dominantly characterize the energy dependence in the
corresponding $\eta_{00}$ functions.
Moreover, Fig.\ref{fig6}(b) demonstrates that the multiple divergences of the $\eta_{00}$ parameter occur around an envelope curve.
This overall behavior can be captured by defining an averaged $\eta_{00}$ according to the relation
\begin{equation}
\braket{\eta_{00}}=\frac{1}{2\delta\nu}\int_{\nu -\delta\nu }^{\nu + \delta\nu}
\eta_{00}(\nu )d\nu
\end{equation}
which will give a smooth curve in the limit that the energy
spacing is smaller than $\delta\nu$.

\subsubsection{Off-center scattering for a square well confining potential}\label{secOCS}

The above results for the infinite square and infinite rectangular wells
are based on the assumption that the scattering center, i.e. the short-range potential
$\hat{V}_s$, is placed at the center of the confining potential.
In this section, the confining potential is considered to
be an infinite square well where the scattering center is located
at a generic point, $(x_s,y_s)=a_\perp (\tilde{x}_s,\tilde{y}_s)$
whereby the transverse  wave function $|\Phi (x_s,y_s)|$ is no
longer only 0 or $2/a_\perp$ as in the previous cases with $(x_s,y_s)=(0,0)$. 
Note that similar systems have been investigated where the interacting
particles are confined in separate harmonic waveguide geometries
yielding in this manner confinement-induced interlayer
molecules \cite{delmerpra2015}.

In the following, the discussion restricted on $s-$wave scattering which couples on even $z$-parity states of the confining square well and/or $p$-wave collisions which couple on odd $z$-parity states of the corresponding confining potential.

In this case the eigenenergies are defined as follows:
\begin{equation}
E_{n_xn_y}=\frac{\hbar^2\pi^2}{2\mu a_\perp^2}(n_x^2 + n_y^2)
\end{equation}
but now the $n_x$ and $n_y$ can independently be $1,2,3$.
The one dimensional wave function is
$\psi (x)=\sqrt{2/a_\perp }\sin (n_x\pi [\tilde{x}-1/2])$.
The transverse eigenfunction of the corresponding confining Hamiltonian $\hat{H}_c$ at the scattering center behave as
\begin{equation}
\Phi_\alpha (x_s,y_s) = \frac{2}{a_\perp}
\sin (\tilde{k}_x [\tilde{x}_s-1/2])
\sin (\tilde{k}_x [\tilde{y}_s-1/2])
\end{equation}
where $\tilde{k}_x=n_x\pi$ and $\tilde{k}_y=n_y\pi$.
The only nonzero terms when the scattering center is at the
origin is from the $n_x$ and $n_y$ which are the odd
integers.

For the case of $s$-wave scattering which couples on even $z-$parity states of the square well and for the case of $p$-wave scattering which couples on odd $z$-parity states the prescription given in the appendix \ref{secGSW} defining in this manner the corresponding $\eta$-parameters.
Namely, the $\eta_{00}$ and $\eta_{10}$ functions fulfill the following relations:
\begin{eqnarray}\label{eqEtaSWOC}
\eta_{00}=&\null&4\sum_{\alpha_0^2 >\epsilon}^\infty
R_{n_x,n_y}(\tilde{x}_s,\tilde{y}_s)
\frac{e^{-2\pi\sqrt{\alpha_0^2-\epsilon}\Delta\tilde{z}}}
{\sqrt{\alpha_0^2-\epsilon}}\cr
&\null&-4\int_{\sqrt{\epsilon}}^\infty
\frac{e^{-2\pi\sqrt{\alpha^2-\epsilon}\Delta\tilde{z}}}
{\sqrt{\alpha^2-\epsilon}}\frac{\pi}{2}\alpha d\alpha\cr
\eta_{10}=-&\null&48\pi^2\sum_{\alpha_0^2 >\epsilon}^\infty
R_{n_x,n_y}(\tilde{x}_s,\tilde{y}_s)
\sqrt{\alpha_0^2-\epsilon}\;e^{-2\pi\sqrt{\alpha_0^2-\epsilon}\Delta\tilde{z}}
\cr
&\null&+48\pi^2\int_{\sqrt{\epsilon}}^\infty
\sqrt{\alpha^2-\epsilon}\;e^{-2\pi\sqrt{\alpha^2-\epsilon}\Delta\tilde{z}}
\frac{\pi}{2}\alpha d\alpha,
\end{eqnarray}
with
\begin{eqnarray}
&\null&R_{n_x,n_y}(\tilde{x}_s,\tilde{y}_s)=\frac{a^2_\perp}{4}
|\Phi_\alpha (x,y)|^2\cr&\null&\null \null \null~~~~~~~~~~~~~~~~~~=
\sin^2 (\tilde{k}_x [\tilde{x}-1/2])
\sin^2 (\tilde{k}_y [\tilde{y}-1/2]).
\end{eqnarray}

where $\alpha_0^2=(n_x/2)^2+(n_y/2)^2$. In the integrals,
the double integrals $d\alpha_xd\alpha_y$ have been converted to cylindrical
coordinates and the integral over angle from $0$ to $\pi /2$ has been
carried out. It may appear that these expressions do not satisfy the
condition that the sum and the respective integral should have the
same form in order for their divergences to cancel and leave a finite difference. To see that the condition still holds, note that the sum
is over 4 times as many terms as the centered square well, {\it but}
the average value of $R_{n_x,n_y}(\tilde{x}_s,\tilde{y}_s)$ is 1/4.
Numerically tests have confirmed that the resulting equations for
$\Gamma$ and $\eta_{00}$,$\eta_{10}$ exhibit
excellent agreement, and they depict features similar to those
shown in Figs.~\ref{fig1} and \ref{fig2}.

\begin{figure}
\includegraphics[scale=0.4]{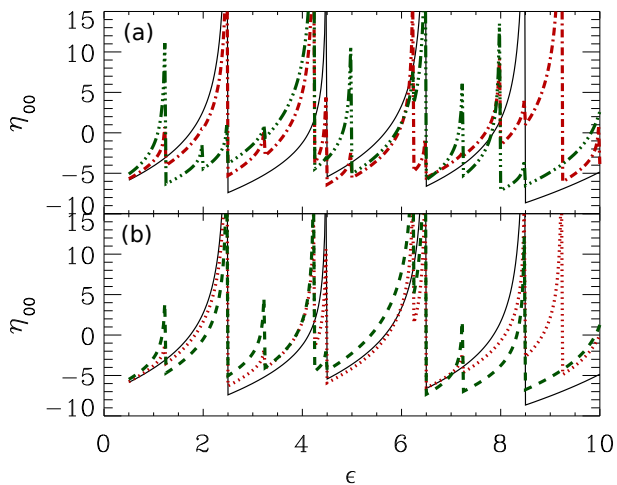}
\caption{(color online) The parameter $\eta_{00}$ for infinite square well confinement
is shown as a function of the scaled energy
$\epsilon = E/(4\hbar^2\pi^2/[2\mu a_\perp^2] )$ with
the scattering center at the scaled position $(\tilde{x}_s,\tilde{y}_s)$:
Panel (a) $(0.0,0.0)$ is the black solid line,  $(0.1,0.1)$ is the red dot-dash line,
and $(0.2,0.2)$ is the green dot-dot-dot-dash line.
Panel (b) $(0.0,0.0)$ is the black solid line, $(0.1,0.0)$ is the red dotted line,
$(0.2,0.0)$ is the green dashed line. The solid line
is the same result as in Fig.\ref{fig4}.} 
\label{fig7}
\end{figure}

Figure \ref{fig7} depicts the $\eta_{00}$ function for different
positions of the scattering center: $(\tilde{x}_s,\tilde{y}_s)$.
As in Fig.~\ref{fig4}, the quantity $\eta_{00}$ is singular as the energy approaches thresholds from below.
In addition, Fig.~\ref{fig7} demonstrates that there are many
more thresholds when the scattering center is shifted from the origin.
This is due to the possibility of scattering into states
that have a node at the origin.
In discussing the position of thresholds, we will only
give the values for $n_x\leq n_y$.
The black solid line in panels (a) and (b) refers to the position $(0,0)$ and is the
same result as plotted in Fig.~\ref{fig4}.
At this position, there are only singularities at the
values of $n_x,n_y$ equal to odd integers.
The thresholds in Fig.~\ref{fig7} are 1,3 at
$\epsilon =2.5$, 3,3 at $\epsilon=4.5$, 1,5 at $\epsilon = 6.5$,
and 3,5 at $\epsilon = 8.5$.
In Fig.\ref{fig7}(b) the $\eta_{00}$ parameter is shown for a scattering center being
placed along the $x$-axis, namely at $(0.1,0.0)$ (red dotted line) and $(0.2,0.0)$
(green dashed line) can have additional singularities than in the case of a scatterer
placed at the origin [see black solid line in Fig.\ref{fig7}(b)].
Specifically, the additional structure emerges when one of $n_x$,$n_y$ is
an even integer.
The additional singularities over those for the scatterer at the center are for $1,2$ at
$\epsilon = 1.25$, for $2,3$ at $\epsilon= 3.25$, for $1,4$
at $\epsilon = 4.25$, for $3,4$ at $\epsilon= 6.25$, for $2,5$
at $\epsilon = 7.25$, and for $1,6$ at $\epsilon =9.25$.
In Fig.\ref{fig7}(a) the $\eta_{00}$ parameter is shown for a scattering center being placed
along the diagonal of the $x-y$ plane, namely at $(0.1,0.1)$ (red dot-dash line) and
$(0.2,0.2)$ (green dot-dot-dot-dash line) where both curves possess additional
threshold singularities in comparison with the black solid line and the curves in Fig.~\ref{fig7}(b).
Specifically, the additional structure emerges when both $n_x,n_y$ are even integers.  
The additional threshold singularities are for $2,2$ at
$\epsilon = 2$, for $2,4$ at $\epsilon =5$, for $4,4$ at
$\epsilon =8$ and for $2,6$ at $\epsilon =10$.

The coefficient of the diverging term is proportional to $R_{n_x,n_y}$ and, thus,
has different effect
depending on the value of the transverse function at the
scattering center that corresponds to that threshold.
For example, the divergence at $\epsilon =2$ (corresponding
to 2,2) is clearly visible when the scattering center
is at $(0.2,0,2)$ but is not visible when the center is at $(0.1,0.1)$.
This is because the latter position is not far shifted
from the origin so that it is still near enough to the node
of the transverse function to be suppressed.
The divergence at $\epsilon = 5$ (corresponding to 2,4)
is visible in both because the larger quantum number increases
the amplitude of the transverse function at the point $(0.1,0.1)$.
Another interesting feature is that the divergence at
$\epsilon = 4.25$ corresponding to $1,4$ has a large strength
partly because it is near the divergence at $\epsilon = 4.5$
corresponding to 3,3.

There are similar types of features in the plots of $\eta_{10}$.
However, because $\eta_{10}$ does not diverge at thresholds,
there are discontinuities in slope at the newly allowed thresholds.

\subsection{Photoabsorption cross section in a harmonic and square well confining geometries}

\begin{figure}
\includegraphics[scale=0.45]{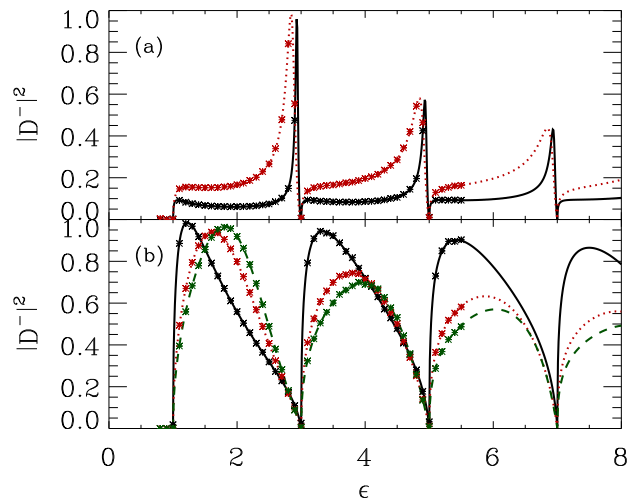}
\caption{(color online) The total absorption probability $|D^-|^2$ (in arbitrary units) for a harmonic oscillator
confining potential as a function of the scaled
energy, $\epsilon =E/(\hbar\omega_\perp )$.
The lines correspond in calculations using Eq.~(\ref{eqPabsTot})
with $(\ell, m)=(0,0)$.
The symbols refer to the ab initio quantum calculations using
coupled angular momenta for the energy range $0.9\leq\epsilon\leq 5.5$.
In panel (a) the $s$-wave scattering lengths take the values of
$-a_\perp /4$ (black solid line) and $-a_\perp /2$ (red dotted line).
In panel (b) the $s$-wave scattering lengths refer to values
of $a_\perp /4$ (black solid line), $a_\perp /2$ (red dotted line), and
$0.68475 a_\perp $ (green dashed line).
}
\label{fig8}
\end{figure}

This subsection investigates the impact of the transversally
harmonic or square well confining potentials on the
photoabsorption cross sections.
In Fig.~\ref{fig8}, the
absorption probability is plotted as a function of the scaled energy
$\epsilon=E/\hbar \omega_\perp$ for the harmonic oscillator confining potential.
The different plots are for different scattering lengths for
$(\ell, m)=(0,0)$.
This figure has both lines and symbols.
The lines are from Eq.~(\ref{eqPabsTot}) while the symbols
are from an ab initio quantum calculation that used a grid
in $r$ and angular momentum to compute the total outgoing flux.
The only adjustment between the two types of calculations
was an overall scale size.
The good agreement between the two types of calculations
shows that Eq.~(\ref{eqPabsTot}) is an excellent approximation.

\begin{figure}
\includegraphics[scale=0.42]{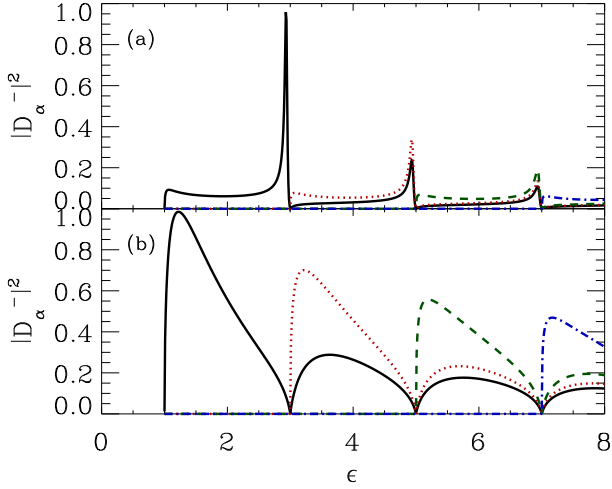}
\caption{(color online) The partial absorption probability, $|D_\alpha^-|^2$ (in arbitrary units)
for a harmonic oscillator confining potential in arbitrary
units as a function of the scaled energy, $\epsilon =E/(\hbar\omega_\perp )$.
The lines are calculations using Eq.~(\ref{eqPabsTot}) with $(\ell, m)=(0,0)$.
Panels (a) and (b) correspond to $s$-wave scattering
lengths of $-a_\perp /4$ and $a_\perp /4$, respectively.
The black solid line is for absorption into the 1st open channel,
red dotted line is 2nd open channel, green dashed line is 3rd open
channel, and blue dot-dashed line is 4th open channel.
}
\label{fig9}
\end{figure}

Two interesting features are worth noting.
The most obvious feature is that the {\it total} probability goes
to 0 at {\it every} threshold.
The probability goes to zero as the threshold is approached from
below and from above.
As the energy approaches a channel threshold from below, the parameter
$\eta_{00}\to \infty$ which means that $|D^-|^2\to 0$ since the
$\eta_{00}$ is in the denominator of $|D^-|^2$.
As the energy approaches a channel threshold from above, 
$N_{00}^2\to\infty$ because the corresponding channel momentum
vanishes, namely $k_\beta\to 0$.
In the expression for $|D^-|^2$, there is an $N_{00}^2$ in the numerator
and an $N_{00}^4$ in the denominator yielding $|D^-|^2\to 0$.
The {\it total} absorption probability is proportional to
$\sqrt{|E-E_\alpha |}$ at each channel threshold $E_\alpha$.
This behavior is in strong contrast to the unconfined absorption
probability where there is a discontinuity in the slope at each
channel threshold, but the total absorption does not go to zero
at each threshold.
Also, in the case of confinement, the photoabsorption
probabilities possess discontinuous slopes near the channel thresholds.

Another important feature is that a negative scattering length causes a
resonance-like structure just below each channel threshold [see Fig.\ref{fig8}(a)].
On the other hand in Fig.\ref{fig8}(b) no resonant structure emerges below each threshold if the $s$-wave scattering length is positive.
This behavior can be understood by the fact that for negative $s$-wave scattering lengths the term $1+a_s(E) \eta_{00}(E)$ in the denominator of the total absorption probability vanishes yielding in this manner a maximum in the total probability.
This effect clearly originates from the closed channel physics.
In addition, in Fig.\ref{fig8}(a) we observe that for smaller (in magnitude) values of scattering length, the
resonance occurs closer to threshold and is narrower in energy.
This is to be expected because the smaller (in magnitude)
scattering length corresponds to a state that is more weakly
bound and, hence, a smaller overlap with the scattering center
leading to a longer lifetime and narrower resonant features.

Figure \ref{fig9} depicts the partial absorption probability $|D_\alpha^-|^2$ for
a negative [see Fig.\ref{fig9}(a)] and a positive $s$-wave scattering length [see Fig.\ref{fig9}(b)] case.
The partial cross section is proportional to $|\mathcal{U}_{\alpha ,00}|^2$
which is proportional to $1/k_\alpha$ since the $\Phi_\alpha (0,0)$
has no dependence on the $\alpha$ channel quantum number.
The channel momentum $k_\alpha$ decreases with increasing
$\alpha$ which means the partial cross section increases
with increasing $\alpha$.
Also, the partial absorption cross section exhibits differences
between channels just above a channel threshold as it is
demonstrated in Fig.~\ref{fig9}, i.e. in panels (a) and (b) for the energy interval $3<\epsilon<5$ the $|D_\alpha^-|^2$ of the second channel (red dotted line) is larger than the partial photoabsorption in the first channel.
In this case, the channel that just opened will have
$k_\alpha\sim 0$ and will therefore have nearly all of the outgoing probability.

\begin{figure}
\includegraphics[scale=0.42]{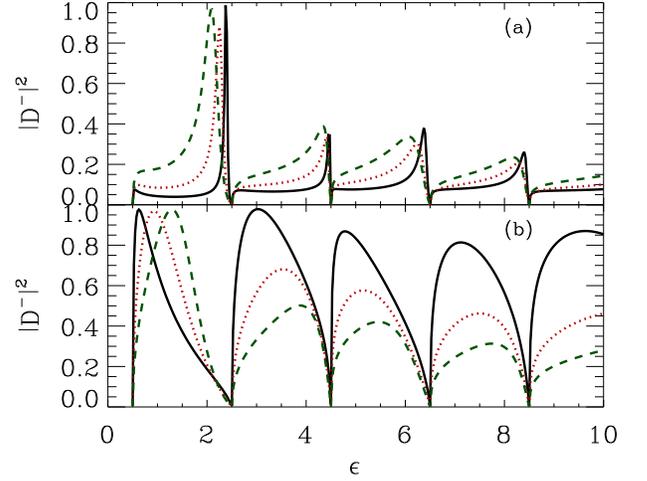}
\caption{(color online) The total absorption probability  $|D^-|^2$ (in arbitrary units) for an infinite square
well confining potential in arbitrary units as a function of
the scaled energy $\epsilon = E/(4\hbar^2\pi^2/[2\mu a_\perp^2] )$
with $(\ell,m)=(0,0)$.
The lines refer to calculations based on Eq.~(\ref{eqPabsTot}).
In panel (a) the $s$-wave scattering lengths acquire the
values $-a_\perp /16$ (black solid line), $-a_\perp /8$ (red dotted line)
and $-a_\perp /4$ (green dashed line).
In panel (b) the $s$-wave scattering lengths have the
values $a_\perp /16$ (black solid line), $a_\perp /8$ (red dotted line)
and $a_\perp /4$ (green dashed line).
}
\label{fig10}
\end{figure}

Figure \ref{fig10} demonstrates the cases of total absorption
probabilities for the infinite square well case for negative
[see Fig.~\ref{fig10}(a)] and positive [see Fig.~\ref{fig10}(b)] $s$-wave
scattering lengths.  
The black solid lines indicate the $s$-wave scattering length with magnitude
$a_\perp /16$, the red dotted lines refer to magnitude $a_\perp /8$
and the green dashed lines denote magnitude $a_\perp /4$.
The general features are the same as for Fig.~\ref{fig8}.
Unlike the Fig.~\ref{fig8} case, there are degenerate channel
thresholds for the square well confining potentials.
For example, the first excited threshold at $\epsilon = 2.5$
is doubly degenerate: $n_x,n_y$ of 0,1 and 1,0.
Also, not all thresholds are equally spaced, in particular for $\epsilon>10$ which leads to
features that change with threshold.

\subsection{Negative-ion photodetachment in magnetic fields}

\begin{figure}
\includegraphics[scale=0.45]{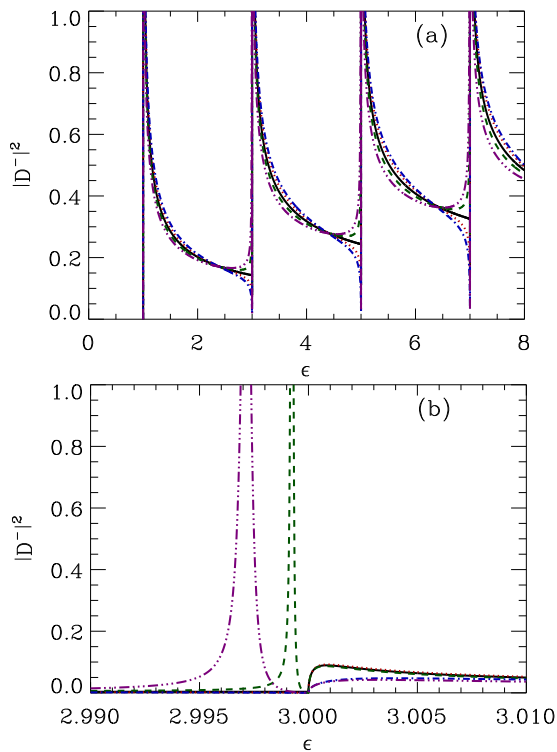}
\caption{(color online) The total absorption probability  $|D^-|^2$ (in arbitrary units) for a harmonic oscillator
confining potential is plotted as a function of the scaled
energy, $\epsilon =E/(\hbar\omega_\perp )$. In panel (a) and (b)
the lines are calculations using Eq.~(\ref{eqPabsTot}).
The black solid line is the result when $\eta_{00}=0$.
The red dotted and green dashed lines correspond to $a_s(E)/a_\perp =0.02$
and $-0.02$ respectively.
The blue dash-dot and purple dot-dot-dash-dot lines are $a_s(E)/a_\perp = 0.04$ and $-0.04$
respectively.
Panel (b) consists of a magnification of panel (a) around the threshold energy $\epsilon= 3$.
}
\label{fig11}
\end{figure}

The result of photodetachment of an electron from a negative
ion in a magnetic field was treated in Ref.~\cite{Greene1987pra}
and is similar
to the example of photodetachment in an isotropic, two-dimensional
harmonic oscillator potential.
However, in Ref.~\cite{Greene1987pra}, only the effect of
open channels is taken into account.
This is equivalent to setting $\eta_{\ell m}(E)=0$ in
Eq.~(\ref{eqPabsTot}) above.
Thus, the results in Ref.~\cite{Greene1987pra} are an approximation
to the full treatment which includes the effects of the closed channels.
Since the formalism of Ref.~\cite{Greene1987pra} has been
successfully applied in many circumstances,
it is worth investigating the regimes where this approximation
($\eta_{\ell m}=0$) is adequate.
Figure \ref{fig11} depicts the total photoabsorption probability
(in arbitrary units) for different $s$-wave scattering
lengths that are a small fraction of $a_\perp$ which is the
typical case for photodetachment in laboratory strength magnetic
fields. The black solid line is the result from the approximation of
Ref.~\cite{Greene1987pra}.
For small positive scattering length (see red dotted and blue dash-dot line in Fig.\ref{fig11}(a)) , the photoabsorption
cross section vanishes below {\it and} above each channel threshold whereas the black solid line vanishes only above every the threshold.
Note that the cross section vanishes around every threshold also for negative scattering lengths [see green dashed and purple dot-dot dashed lines in Fig.\ref{fig11}(a)].
Moreover, the case of negative scattering lengths exhibits another qualitative difference from the approximation of Ref.~\cite{Greene1987pra}.
More specifically, this can be seen in Fig.\ref{fig11}(b) where the total photoabsorption probability for negative scattering length (green dashed and purple dot-dot-dashed lines) possesses a resonance barely below the channel threshold. This non-trivial resonant feature is absent from the black solid line manifesting in this manner the importance of the closed channel physics.
In addition, Fig.\ref{fig11} (a) and (b) shows that away from the channel thresholds our results are practically the same as in the approximation of Ref.~\cite{Greene1987pra}, [black solid line in Fig.\ref{fig11}].
Note that there is one energy between each threshold where all of the results are the same.
These are the energies where $\eta_{00}=0$.

Evidently, the role of the closed channels becomes important only around each threshold due to small scattering lengths.
This explains why the result of Ref.~\cite{Greene1987pra} works well for photodetachment in a magnetic field.
In this situation, the scattering length is of the order of a few
Bohr radii while the confinement length is
$a_\perp = a_0\sqrt{B_{a.u.}/B}$ where $a_0$ is the Bohr radius
and $B_{a.u.}$ is the atomic unit of magnetic field.
Thus, even in a 1~T magnetic field, the scattering length will
typically be less than $a_\perp /100$.
However, we expect that experiments would show differences
from the treatment of Ref.~\cite{Greene1987pra} in stronger
magnetic fields or for a negative ion whose photodetachment probes a final state having a much larger electron-atom scattering length.

\begin{figure}
 \includegraphics[width=0.45\textwidth]{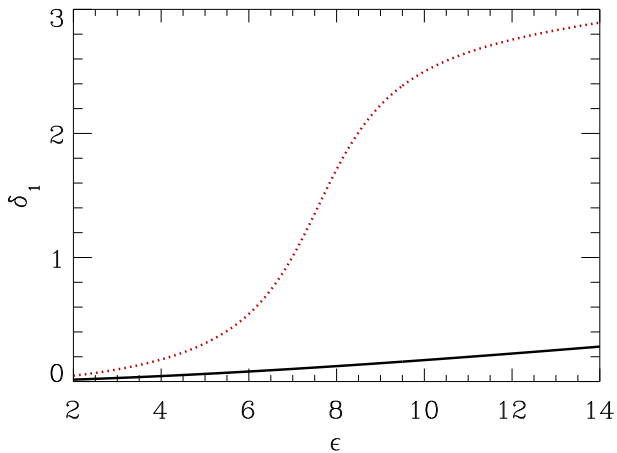}
\caption{(color online) The $\ell =1$ phase shifts used in the calculations of
Fig.~\ref{fig13}. The black solid line is a non-resonance example and
the red dotted line is the resonance case.
The energy of the resonance is $\epsilon_r=7.8$.}
\label{fig12}
\end{figure}

\begin{figure}[h!]
\includegraphics[scale=0.8]{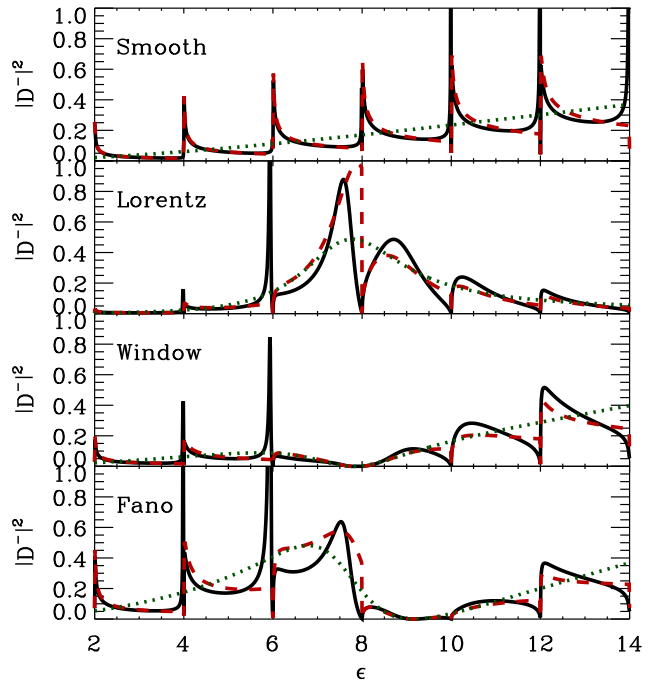}
\caption{(color online) The $\ell =1$ photoabsorption  $|D^-|^2$ (in arbitrary units) for different cases.
In all calculations, the green dotted line is for photoabsorption with no
confining potential, the red dashed line is the approximation that includes
the confining potential but does not include the effect of closed
channels (equivalent to setting $\eta_{11}=0$), and the black solid line
is the full calculation including the confining potential.
The ``Smooth" calculation does not have a resonance.
The ``Lorentz", ``Window", and ``Fano" calculations only differ in
the energy dependence of the dipole matrix element when no field is applied.
}
\label{fig13}
\end{figure}

The photodetachment for higher partial waves may possess shape
resonances which allows for a larger effect from
the closed channel collective state described by $\eta_{\ell m}(E)$.
Therefore in the following, consider a photodetachment process
into $\ell =1,m=1$ for a harmonic oscillator confinement.
The confining potential does not have as strong an effect on
the $\ell =1, m=0$ case because the outgoing flux is mostly
along the $z$-axis compared to $m=1$ where the outgoing flux is
mainly perpendicular to the $z$-axis.
The energy dependent scattering volume is given by the relation
$\tan\delta_1(E) /[k a_\perp]^3$ where a resonance form for the
phase shift, $\delta_1$ is chosen according to the relation:
\begin{equation}
\tan\delta_1 = -k^3 \left[f_{sc}a_\perp^3 +
\frac{\gamma /k_r^3}{\epsilon -\epsilon_r}\right]
\end{equation}
where $\epsilon =E/(\hbar\omega_\perp )$ is the scaled energy
with $E=\hbar^2k^2/2\mu$, $\epsilon_r=E_r/(\hbar\omega_\perp )$
is the scaled energy of the resonance with $E_r=\hbar^2k_r^2/2\mu$,
$\gamma$ is a parameter proportional to the scaled energy width of
the resonance, and $f_{sc}$ is the scaled scattering volume.
Figure~\ref{fig12} shows the $\delta_1$ phase shifts for a resonance
case (red dotted line) with $f_{sc}=-(1/8)^3$, $\gamma =1.4$, and
$\epsilon_r=7.8$ and for a non-resonance case (black solid line) $f_{sc}=-(1/8)^3$ and $\gamma = 0$.
For the energy dependent smooth dipole, $D_s(E)$, the following form is chosen
\begin{equation}
D_s(E) = k^{3/2}\left[d_s + (d_r/a_\perp^3)
\frac{\gamma /k_r^3}{\epsilon -\epsilon_r}\right]
\end{equation}
where $d_s$ represents the amplitude for photoabsorption directly
to the continuum and $d_r$ represents the amplitude for
photoabsorption into the resonance.
Figure~\ref{fig13} shows the results from 4 different calculations.
In all calculations, the green dotted line is the photoabsorption that
would occur if there were no confining potential, the black solid line
corresponds to the full calculation with the confining potential, and
the red dashed line is the approximation in Ref.~\cite{Greene1987pra}
where the effect of the closed channels is neglected which is
equivalent to setting $\eta_{11}=0$.
As expected, the resonance features from the confinement calculations
give an overall shape that tracks the non-confinement calculations,
but with sharp
features near the thresholds.
The topmost calculations are when there is no resonance $\gamma =0$.
The ``Lorentz" calculation has $d_s=0$; it does not look like the
standard Lorentzian shape resonance due to the overall increase
proportional to $\epsilon^{3/2}$.
The ``Window" calculation has $d_r=0$; as with the Lorentz calculation,
this does not look like the symmetrical window resonance due to the
overall increase.
The ``Fano" calculation has $d_r/d_s=60$ which gives a 0 in the
dipole matrix element at $\epsilon\simeq 9.1$ which is at a
somewhat higher energy than the resonance position at 7.8.
The full and approximate confinement calculations differ most
strongly where $\tan\delta_1$ is large which is near the resonance
{\it and} an energy range near the thresholds. For the three cases
with the resonance at $\epsilon_r=7.8$, the absorption cross
section shows sharp resonances below each threshold only for
$\epsilon < \epsilon_r$ because the scattering volume is
negative in this range.

\section{Summary and Conclusions}

\subsection{Summary of Method}

It is evident that there are many steps in the derivation and application of the Schwinger variational approach to collisions involving a short range spherically symmetric potential in the presence of various types of confining geometries.
Therefore, in this section, the important steps are pointed out for applying the method. First, Eq.~(\ref{allklm})
gives the expression for the $K$-matrix and Eq.~(\ref{eqDalph}) gives
the dipole matrix element in terms of the scattering length or volume
and the energy dependent parameters $\mathcal{U}_{\alpha ,\ell m}$ and
$\eta_{\ell m}$. The conversion from spherical coordinates to the
confinement geometry is encapsulated in the $\mathcal{U}_{\alpha ,\ell m}$ in
Eq.~(\ref{lft}) which depends on the confinement wave functions
$\Phi_\alpha (x,y)$ at the scatterer.

In this paper, the expressions for the $\eta_{\ell m}$ were obtained
from a detailed examination of the confinement Green's function, $\hat{G}_c$, and
various integral expressions for the free Green's function, $\hat{G}_f$.
There is a conceptual shortcut that can be used to obtain the
$\eta_{\ell m}$: the $\hat{G}_c$ becomes the $\hat{G}_f$ in the limit that the
confinement length scale, $a_\perp$, goes to infinity. This is the
reason that the $\eta_{\ell m}$ have the form of a sum (from the $\hat{G}_c$)
minus an integral that has the same functional form
(from the $\hat{G}_f$ as formulated from $\hat{G}_c$ in the limit
$a_\perp\to\infty$).
Similar to Eq.~(\ref{lft}), define a transformation for the closed
channels
\begin{equation}\label{lftcl}
\bar{\mathcal{U}}_{\alpha,\ell m}= \begin{cases}
		  \sqrt{\frac{2\pi }{\tilde{\kappa }_\alpha}}
		   a_\perp\Phi_\alpha (\boldsymbol{\rho}_s) & \ell=m=0 \\
		  \sqrt{6\pi \tilde{\kappa }_\alpha}
		  a_\perp\Phi_\alpha (\boldsymbol{\rho}_s) & \ell=1,~m=0 \\
		  -\sqrt{\frac{12\pi }{\tilde{\kappa }_\alpha}}a_\perp^2 (\partial_{\pm}\Phi_\alpha )\big|_{\boldsymbol{\rho}_s} & \ell=1,~m=\pm1
		 \end{cases}
\end{equation}
where the operators are defined below Eq.~(\ref{eqNump1}),
$\boldsymbol{\rho}_s$ is the position of the scatterer, and
$E=-\hbar^2\tilde{\kappa }^2_\alpha /(2\mu a_\perp^2 ) + E_\alpha$.
Since $\Phi_\alpha$ is proportional to $1/a_\perp$, the $\bar{\mathcal{U}}$ only
depends on the form of the confining potential, the {\it scaled} energy, and
the $\boldsymbol{\rho}_s/a_\perp$. In this paper,
we restricted the confining potential to be independent of $z$ and
symmetric in $xy$. For these cases, the following relation between
$\eta$ and $\bar{\mathcal{U}}$ hold
\begin{eqnarray}\label{lfteta}
\eta_{\ell m}=(-1)^{\ell+m} &\null &[\sum_{\alpha >\alpha_o}^\infty
|\bar{\mathcal{U}}_{\alpha ,\ell m}|^2
e^{-\tilde{\kappa}_\alpha \Delta\tilde{z}}\cr
&-&\int_{\alpha \geq \alpha_o}^\infty |\bar{\mathcal{U}}_{\alpha ,\ell m}|^2
e^{-\tilde{\kappa}_\alpha \Delta\tilde{z}}d\alpha ]
\end{eqnarray}
where the limit $\Delta\tilde{z}\to 0^+$ is understood. The
integral over $d\alpha $ is the result of the quantum numbers for
$G_f$ giving an infinitesimal spacing of $\kappa_\alpha$ in the
limit $a_\perp\to\infty$. For the case of Sec.~\ref{secOCS} which
does not satisfy the restrictions on the confining potential, the
$|\bar{\mathcal{U}}_{\alpha ,\ell m}|^2$ is replaced with the average value
as $a_\perp\to\infty$.

\subsection{Conclusions}

In the preceding sections, the development of a Schwinger variational
framework is presented as a treatment of scattering in
a confined geometry.
The current formalism is non-perturbative, permitting the treatment of a
wide class of Hamiltonian systems which possess two potentials.
The particular type of Hamiltonians studied in this work consists
of systems where the interaction at short distances is spherically
symmetric and, at large distances, a confining potential is considered
which bounds the motion of the particle in the degrees of
freedom perpendicular to the direction of its propagation.
The restriction in the current formalism amounts to the fact
that the considered potentials dominate at different scales. For
the sake of simplicity, we
restricted the confining potential to have no dependence in the
$z$-coordinate and symmetry in $xy$. However, this theoretical
method could treat general confining potentials where length
scale separation holds.

The theoretical formalism presented in this paper allows a self-consistent
treatment of scattering in a confined geometry where the results
are manifestly convergent at each step of the derivation. This formalism
has been used to derive results obtained in previous studies
(e.g. s- and p-wave scattering within a harmonic confining potential
and s-wave scattering in a square well confining potential) and
to quickly derive results in novel geometries (e.g. s-wave and p-wave
scattering in off-center confining potentials and scattering in
rectangular confining potentials). Results for the scattering parameters
in different geometries were presented and main qualitative features
were explained. The formalism was also used to derive
a treatment of half-scattering problems (e.g. photo-detachment) in
confining potentials. For example, the photo-detachment results in
a harmonic confinement was compared
to a previous LFT treatment that only accounted for open channels to
assess the role of the closed channels which are included in this
treatment: the effect from the closed channels are most important
near thresholds and become more important as the scattering length
becomes a sizable fraction of the confinement length scale.

An interesting possibility for the formalism in this paper is to
aid in the understanding of the implementation and possible limitations
of the LFT. The basic steps in the current formalism were cast
in a form strongly reminiscent of the LFT.
However, these two frameworks possess a conceptual difference
which is mainly focused on the physics of the closed channels.
More specifically, in the LFT approach the $K$-matrix formulas
include diverging sums and therefore regularization
schemes are employed to remove such singularities.
In the Schwinger variational approach, such
divergences do not emerge, which eliminates the need for tricky  regularization techniques that are often used in the LFT theory.
In particular, we observe that in the physical $K$-matrix of
the LFT the information of the free space $\hat{G}_f$ is absent
and it has been indirectly incorporated via regularization schemes \cite{GrangerBlume2004prl,Giannakeas2012pra,Zhang2013pra}.

In view of the rigorousness of the Schwinger variational
approach, we expect that it can be equally applied
to other physical systems where the LFT method has been used.
For example, this method can be used to derive the scattering
of an electron in a Rydberg state from a neutral perturber and,
for $\ell =0$, the same result as Eq.~(14) of Ref.~\cite{dupra1987}
is obtained.
Note that the frame transformation ideas have been used in many
different circumstances.
A short list of examples could include the $jj-LS$ transformation,
molecular rotational transformations, molecular vibrational
transformations, local frame transformations involving external
electric and/or magnetic fields.
In all cases, wave functions in one representation are projected
on those in another representation at a surface (or indirectly in a region) where both
representations are expected to be accurate.
However, the level of error involved in such a procedure is not clear.
It is conceivable that the method described above might enable a
more systematic derivation of the frame transformation so
that the level of expected error would be clearer.

\begin{acknowledgments}
This work was supported by the U.S. Department of Energy, Office
of Science, Basic Energy Sciences, under Award numbers DE-SC0010545
(for PG and CHG) and DE-SC0012193 (for FR).
\end{acknowledgments}

\appendix

\section{Matrix elements with $\hat{V}_s\ket{\psi_s}$}\label{secMEVPapp}

The volume integral in Eq.~(\ref{eqPsisC}) can be simplified
by employing the relation $\hat{V}_s\ket{\psi_s} = (E-\hat{H}_f)\ket{\psi_s}$.
Then, assuming length scale separation, the confining potential
$\hat{V}_c$ is practically zero in the range of $\hat{V}_s$ where the
following relation is valid
$(E-\hat{H}_c)\ket{\psi_{c}}\approx(E-\hat{H}_f)\ket{\psi_{c}}\approx0$.
Thus, integrating by parts twice Eq.~(\ref{eqPsisC}), the terms
of $(E-\hat{H}_f)\ket{\psi_{c}}$ vanish and only the surface ones survive.
In general, the following relation is fulfilled:
\begin{equation}
\braket{\psi_c|\hat{V}_s|\psi_s}=\frac{\hbar^2}{2\mu}
\int_{\sigma(\tau)} \hat{n}\cdot \{\psi_c^*(\mathbf{r})[\vec{\nabla}\psi_s(\mathbf{r})]
-[\vec{\nabla} \psi_c^*(\mathbf{r})]\psi_s(\mathbf{r})\}d\sigma
\end{equation}
where $\sigma(\tau)$ indicates the surface containing the
volume $\tau$ and $\hat{n}$ is the unit vector which is
outward normal to the surface.
If the region $\tau$ is a sphere of radius $r$, this matrix element
can be written as
\begin{equation}\label{eqWron}
\braket{\psi_c|\hat{V}_s|\psi_s}=\frac{\hbar^2}{2\mu}r^2
\int \bigg[\psi_c^*(\mathbf{r})\frac{\partial\psi_s(\mathbf{r})}{\partial r}
-\frac{\partial \psi_c^*(\mathbf{r})}{\partial r}\psi_s(\mathbf{r})\bigg]d\Omega
\end{equation}
where $d\Omega = \sin\theta d\theta d\phi$ is the solid angle
differential element.

This subsection recast the
integrals which involve $\hat{V}_s\ket{\psi_s}$ into surface integrals
when length scale separation between $\hat{V}_s$ and $\hat{V}_c$ is
a good approximation.
The following subsections focus on special
cases where $\ket{\psi_s}$ is an eigenstate of angular momentum $\ell$
and its projection,
providing explicit analytical equations for this situation.

\subsection{Analytic expressions of $\braket{\psi_c|\hat{V}_s|\psi_s}$ for $\ell = 0,~m=0$}

Assume that in this case study the states
$\ket{\psi_s}\equiv\ket{\psi_{s,\ell m}}$ possess $\ell = 0,~m=0$ character.
At distances where the $\hat{V}_c$ potential is considered small,
the $\psi_c(\mathbf{r})$ to a good approximation can be written
as a linear combination of spherical Bessel functions times
spherical harmonics, $j_\ell (kr)Y_{\ell m}(\Omega )$.
Note that the wave number $k$ corresponds to the total energy
and $\Omega$ denotes the $(\theta,~ \phi)$-angles.
The coefficients of the linear combination can be specified by
expanding in Taylor series the $\psi_c(\mathbf{r})$ and
$j_\ell (kr)Y_{\ell m}(\Omega )$ functions and matching them term by term.
However, at small distances only few terms are needed to be matched.

The coefficient of the $j_0(kr)Y_{00}(\Omega)$ term is
simply $\sqrt{4\pi}$ times the value of $\psi_c(\mathbf{r})$
at the origin giving
\begin{equation}
\psi_c(\mathbf{r})\simeq\psi_c(0)\sqrt{4\pi}j_0(kr)Y_{00}(\Omega )
\end{equation}
After substituting this expression into Eq.~(\ref{eqWron}) the
surface integral yields the following general form:
\begin{equation}\label{eqFmatels}
\braket{ \psi_c|\hat{V}_s|\psi_s}=\sqrt{\frac{\pi\hbar^2}{\mu}}
\psi_c(0)\sqrt{\frac{2\pi}{k }}\left( -\frac{1}{\pi}\tan\delta_0\right),
\end{equation}
where the expressions have been used for the $\psi_c(\mathbf{r}) \equiv \psi^{(e)}_{c,\alpha}(\mathbf{r})$ [see Eq.~(\ref{eqPsi0eo})] and $\psi_s(\mathbf{r})\equiv \psi_{s,00}(\mathbf{r})$ Eq.~(\ref{eqFmatels}) reads
\begin{equation}\label{eqNumsapp}
\braket{\psi^{(e)}_{c,\alpha}|\hat{V}_s|\psi_{s,00}} =
\Phi^*_\alpha (0,0)\sqrt{\frac{2\pi}{k k_\alpha}}\left( -\frac{1}{\pi}\tan\delta_0\right)
\end{equation}

\subsection{Analytic expressions of $\braket{\psi_c|\hat{V}_s|\psi_s}$ for $\ell = 1,~m=0$}

This subsection focuses on the case where
$\ket{\psi_s}\equiv \ket{\psi_{s,\ell m}}$ possesses an
$\ell = 1,~m=0$ character.
In a similar way as was discussed in the previous subsection the
$\psi_c(\mathbf{r})$ functions are matched to the superposition
of spherical Bessel functions times spherical harmonics,
$j_\ell (kr)Y_{\ell m}(\Omega )$.
In this particular case the coefficient of the $j_1(kr)Y_{10}$
term is simply $\sqrt{12\pi}/k$ times the value of
$\partial \psi_c(\mathbf{r})/\partial z$ at the origin.
\begin{equation}
\psi_c(\mathbf{r})\simeq\frac{\partial \psi_c(\mathbf{r})}{\partial z}\Bigg|_{|\mathbf{r}|=0}\frac{\sqrt{12\pi}}{k}j_1(kr)Y_{10}(\Omega )
\end{equation}

Substituting this expression into Eq.~(\ref{eqWron}) the
surface integral reads:
\begin{equation}\label{eqFmatelp}
\braket{\psi_c|\hat{V}_s|\psi_s}=\sqrt{\frac{\pi\hbar^2}{\mu}}
\frac{\partial \psi_c(\mathbf{r})}{\partial z}\Bigg|_{|\mathbf{r}|=0}\sqrt{\frac{6\pi}{k^3 }}
\left( -\frac{1}{\pi}\tan\delta_1\right),
\end{equation}
where using the expressions for the
$\psi_c(\mathbf{r})\equiv\psi^{(o)}_{c,\alpha}(\mathbf{r})$ from Eq.~(\ref{eqPsi0eo}) and $\psi_s(\mathbf{r})\equiv \psi_{s,10}(\mathbf{r})$ Eq.~(\ref{eqFmatelp}) reads
\begin{equation}\label{eqNumpapp}
\braket{\psi^{(o)}_{c,\alpha}|\hat{V}_s|\psi_{s,10}} =
\Phi^*_\alpha (0,0)\sqrt{\frac{6\pi  k_\alpha}{k^3}}\left( -\frac{1}{\pi}\tan\delta_1\right)
\end{equation}

\subsection{Analytic expressions of $\braket{\psi_c|\hat{V}_s|\psi_s}$ for $\ell =1,~m=\pm 1$}

This subsection considers the case where
$\ket{\psi_s}\equiv\ket{\psi_{s,\ell m}}$ possesses
$\ell = 1,~m=\pm1$ character.
Therefore in this case the coefficient of the $j_1(kr)Y_{1\pm1}(\Omega)$
term is simply $-\sqrt{24\pi}/k$ times the value of
$(\partial \psi_c(\mathbf{r})/\partial x \mp i\partial \psi_c(\mathbf{r})/\partial y )/2$ at the origin yielding the following relation:
\begin{equation}
\psi_c(\mathbf{r})\simeq -(\partial_\pm \psi_c)\Bigg|_{|\mathbf{r}|=0}\frac{\sqrt{24\pi}}{k}j_1(kr)Y_{1\pm1}(\Omega )
\end{equation}
where $(\partial_\pm \psi_c)\equiv(\partial \psi_c(\mathbf{r})/\partial x\mp i\partial \psi_c(\mathbf{r})/\partial y )/2$.

Substituting these expressions into the surface integral gives
the simple result:
\begin{equation}\label{eqFmatelp1}
\braket{\psi_c|\hat{V}_s|\psi_s}=\sqrt{\frac{\pi\hbar^2}{\mu}}
(\partial_\pm \psi_c)^*
\Bigg|_{0}\sqrt{\frac{12\pi}{k^3 }}
\left( -\frac{1}{\pi}\tan\delta_1\right),
\end{equation}
where using the expressions for the $\psi_c(\mathbf{r}) \equiv\psi^{(e)}_{c,\alpha}(\mathbf{r})$ from Eq.~(\ref{eqPsi0eo}) and $\psi_s(\mathbf{r})\equiv \psi_{s,1\pm 1}(\mathbf{r})$ Eq.~(\ref{eqFmatelp1}) reads
\begin{equation}\label{eqNump1app}
\braket{\psi^{(e)}_{c,\alpha}|\hat{V}_s|\psi_{s,1\pm1}} =
-(\partial_\pm \Phi_\alpha )^*
 \Bigg|_{0}\sqrt{\frac{12\pi}{ k_\alpha k^3}}\left( -\frac{1}{\pi}\tan\delta_1\right)
\end{equation}

\section{Evaluation of the $D_s$ matrix elements}\label{secDsapp}

At this point, all of the terms in the expression for the
$K$-matrix, Eq.~(\ref{eqSVPs}) have a simple analytic expression
except for the term involving the matrix element of
$\hat{V}_s\Delta \hat{G} \hat{V}_s$, see Eq.~(\ref{eqDelG}).
The most straightforward method for evaluating this matrix element
is to individually compute the matrix elements of
$\hat{V}_s\hat{G}_f\hat{V}_s$ and $\hat{V}_s\hat{G}_c\hat{V}_s$
and subtract.
However, this method is fraught with difficulties because each
of these matrix elements involve integrands that diverge.

At the symmetry point of $\hat{V}_c$, the function
$\Delta \hat{G}$ only has even powers of $(z_1-z_2)$
and powers of $\boldsymbol{\rho}_1\cdot\boldsymbol{\rho}_2$ through
second order if the confining potential is a power series in $x,y,z$.
The vector $\boldsymbol{\rho}_1\equiv\hat{x}x_1+\hat{y}y_1$.
Thus, the matrix element in Eq.~(\ref{eqDelG}) only has terms
from regular functions at least to order $r^3$.
If the state $\ket{\psi_s}$ possesses $\ell =0$ or $\ell = 1,~m$
character, then only terms in $\Delta \hat{G}$ proportional to
$(z_1-z_2)^0$, $(z_1-z_2)^2$, and
$(\boldsymbol{\rho}_1-\boldsymbol{\rho}_2)^2$ will contribute to the integral:
\begin{equation}\label{eqdelGdelz}
\Delta G(\mathbf{r}_1,\mathbf{r}_2) \simeq \Delta G(0) - \Delta G_{zz}(0) z_1z_2
-\Delta G_{\rho\rho}(0) \boldsymbol{\rho}_1 \cdot \boldsymbol{\rho}_2,
\end{equation}
where $\Delta G(0)$, $\Delta G_{zz}$ and $\Delta G_{\rho\rho}$ are
coefficients that will be determined below.
Note that terms proportional to $x_j^2$, $y_j^2$, and $z_j^2$ are
not included since they only affect matrix elements for $d$-wave scattering
or give small corrections to the $s$-wave scattering.
Recall that $\Delta \hat{G}$ is a real function since its
constituents (i.e. $\hat{G}_f$ and $\hat{G}_c$) are the principal
value Green's functions.
Using this expansion, the dominant contribution to the $D_s$ matrix
element for the $\ell =0,~m=0$ case is the $\Delta G(0)$ term
\begin{eqnarray}\label{ds00}
D_{s,00}&=&\braket{\psi_{s,00}|\hat{V}_s|z_1^0} \Delta G(0) \braket{z_2^0|\hat{V}_s|\psi_{s,00}}\cr
&=&\Delta G(0)\frac{\pi\hbar^2}{\mu}\frac{2\pi}{k}\left(-\frac{1}{\pi}\tan\delta_0\right)^2
\end{eqnarray}

For the case of $\ell=1,~m=0$ the dominant contribution to the
$D_s$ matrix elements is from the term $-z_1z_2\Delta G_{zz}(0)$
\begin{eqnarray}\label{ds10}
D_{s,10}&=&-\braket{\psi_{s,10}|\hat{V}_s|z_1} \Delta G_{zz}(0) \braket{z_2|\hat{V}_s|\psi_{s,10}}\cr
&=&-\Delta G_{zz}(0)\frac{\pi\hbar^2}{\mu}\frac{6\pi}{k^3}\left(-\frac{1}{\pi}\tan\delta_1\right)^2
\end{eqnarray}

Similarly, for the case of $\ell=1, m= \pm 1$ the dominant
contribution to the $D_s$ matrix elements emerges from the term
$-\boldsymbol{\rho}_1\cdot\boldsymbol{\rho}_2\Delta G_{\rho\rho}(0)$
\begin{eqnarray}\label{ds11}
D_{s,1 \pm1}&=&-\braket{\psi_{s,1\pm1}|\hat{V}_s|\rho_1e^{\pm i\phi_1}} \Delta \hat{G}_{\rho\rho}(0) \braket{\rho_2e^{\pm i\phi_2}|\hat{V}_s|\psi_{s,1\pm 1}}\cr
&=&-\Delta G_{\rho\rho}(0)\frac{\pi\hbar^2}{\mu}\frac{6\pi}{k^3}\left(-\frac{1}{\pi}\tan\delta_1\right)^2
\end{eqnarray}

\section{$\eta_{\ell m} (E)$ functions for given confining geometries}\label{secEtaapp}

In the expressions for the Green's functions, we only need terms through
$\boldsymbol{\rho}_1\cdot\boldsymbol{\rho}_2$.
The free space Green's function, $\hat{G}_f$ is Taylor
expanded up to terms proportional to
$\boldsymbol{\rho}_1\cdot\boldsymbol{\rho}_2$ which give
\begin{equation}\label{eqGfexp}
\frac{\cos (\tilde{k}\tilde{r}_{12})}{|\tilde{r}_{12}|}
\simeq \frac{\cos (\tilde{k}\Delta\tilde{z})}{|\Delta\tilde{z}|}
+\frac{\boldsymbol{\rho}_1\cdot\boldsymbol{\rho}_2}{a_\perp^2}
\left(\frac{\cos (\tilde{k}\Delta\tilde{z})}{|\Delta\tilde{z}|^3}+
\frac{\tilde{k}\sin (\tilde{k}\Delta\tilde{z})}{\Delta\tilde{z}^2}
\right)
\end{equation}
where the scaled variables are $\tilde{k}=k a_\perp$ and $\Delta\tilde{z}
=\Delta z/a_\perp$. The strategy below will be to write this expression
for $\hat{G}_f$ as an integral with a uniformly converging integrand. We will
show that this integral will be the continuum form of the sums that arise
in $\hat{G}_c$.

In this subsection, the functions
$\eta_{\ell m}(E)$ are derived for two types of confining
potentials, i.e. harmonic and infinitely square well confining potential.
In addition, we compare the approximation Eq.~(\ref{eqGammaapp})
to the full expression for $\Gamma(E)$ as a function of
$\Delta\tilde{z}$, Eq.~(\ref{eqGamma}) as an illustration of the
accuracy of the $\eta_{\ell m}(E)$.

\subsection{Harmonic Oscillator Confining Potential}\label{secGHO}

In this section, the motion of the particle in the transversal
degrees of freedom is bounded by a harmonic oscillator which
possesses the form $V_c(\rho)=\frac{\mu}{2} \omega_\perp \rho^2$
where $\mu$ denotes the mass of the particle, $\omega_\perp$
corresponds to the frequency of the harmonic confinement
corresponding to the length scale $a_\perp= \sqrt{\hbar /\mu \omega_\perp}$.
Note that the variable $\rho$ represents the polar coordinate,
namely $\rho=\sqrt{x^2+y^2}$.

Under these considerations, the motion of the photoelectron
separates in the $z,~\rho$ degrees of freedom.
The solutions of the corresponding Hamiltonian, namely
$\hat{H}_c$, possess the form given in Eq.~(\ref{eqPsi0eo}) for even or odd parity in $z$-direction.
Accordingly, the eigenenergies of the Hamiltonian $\hat{H}_c$
fulfill the relation
$E=\hbar \omega_\perp(2 n +|m|+1)+\hbar^2k^2_{n,|m|}/(2\mu)$,
where $n=0,1,\ldots$.
Evaluating the harmonic oscillator eigensolutions at
the $(\rho,~\phi)\to 0$ the following relations are obtained
\begin{equation}\label{harmoscphi}
 \begin{cases}
		  \Phi_{n,m}(0)=\frac{1}{a_\perp \sqrt{\pi}} & \rm{for}~m=0 \\
		  (\partial_{\pm}\Phi_{n,m})(0)=\frac{\sqrt{n +1}}{a^2_\perp \sqrt{\pi}}& \rm{for}~m=\pm1
 \end{cases}
\end{equation}

The scaled energy is defined by the relation $\epsilon
=E/(\hbar\omega_\perp )$. The scaled momenta can be written as
$\tilde{k}=\sqrt{2\epsilon }$, $\tilde{k}_{nm}=\sqrt{2\epsilon - 2(2n+|m|+1)}$,
and $\tilde{\kappa}_{nm}=\sqrt{2(2n+|m|+1)-2\epsilon}$.
Substituting Eq.~(\ref{harmoscphi}) in Eq.~(\ref{eqGamma}) with the
approximation of Eq.~(\ref{eqGfexp}), the
following relation is obtained
\begin{eqnarray}\label{eqGamHO}
\Gamma(\boldsymbol{r}_1,\boldsymbol{r}_2,E) =
&-&2\sum_{n=0}^{n_{o,0}}\frac{\sin (\tilde{k}_{n0}
\Delta\tilde{z})}{\tilde{k}_{n0}}
+2\sum_{n=n_{o,0}+1}^\infty\frac{ e^{-\tilde{\kappa}_{n0}
\Delta\tilde{z}}}{\tilde{\kappa}_{n0}}\cr
&+&2\int_{-1/2}^{n_{o,0}}\frac{\sin (\tilde{k}_{n0}
\Delta\tilde{z})}{\tilde{k}_{n0}} dn
-2\int_{n_{o,0}}^\infty \frac{ e^{-\tilde{\kappa}_{n0}\Delta\tilde{z}}}{
\tilde{\kappa}_{n0}}dn\cr
&-&\tilde{\boldsymbol{\rho}}_1\cdot\tilde{\boldsymbol{\rho}}_2
4\sum_{n=0}^{n_{o,1}}\frac{(n+1)\sin (\tilde{k}_{n1}\Delta\tilde{z})}{
\tilde{k}_{n1}}
\cr
&+&\tilde{\boldsymbol{\rho}}_1\cdot\tilde{\boldsymbol{\rho}}_2
4\int_{-1}^{n_{o,1}}\frac{(n+1)\sin (\tilde{k}_{n1}\Delta\tilde{z})}{
\tilde{k}_{n1}}dn
\cr
&+&\tilde{\boldsymbol{\rho}}_1\cdot\tilde{\boldsymbol{\rho}}_2
4\sum_{n=n_{o,1}+1}^\infty
\frac{(n+1)e^{-\tilde{\kappa}_{n1}\Delta\tilde{z}}}{\tilde{\kappa}_{n1}}\cr
&-&\tilde{\boldsymbol{\rho}}_1\cdot\tilde{\boldsymbol{\rho}}_2
4\int_{n_{o,1}}^\infty
\frac{(n+1)e^{-\tilde{\kappa}_{n1}\Delta\tilde{z}}}{\tilde{\kappa}_{n1}}dn
\end{eqnarray}
where $n_{o,m}=(\epsilon-1-|m|)/2$ is the divide between open and closed
channels; in the sums, the integer part should be used.
Note that $\tilde{\boldsymbol{\rho}}_1 \cdot \tilde{\boldsymbol{\rho}}_2=\boldsymbol{\rho}_1 \cdot \boldsymbol{\rho}_2/a_\perp^2$.
This expression has many terms but each can be identified with different
parts of the Eq.~(\ref{eqGamma}). The first line is the contribution
from the $m=0$ open (first term) and closed (second term) channels
from $G_c$. The second line is an integral expression that exactly
equals $-\cos(\tilde{k}\Delta\tilde{z})/\Delta\tilde{z}$
(see Eq.~(\ref{eqGfexp})) which is the
contribution to $m=0$ from $\hat{G}_f$. The third (fifth) line is the contribution
to $m=\pm 1$ from the open (closed) channels in $\hat{G}_c$. The fourth
and sixth lines are an integral expression whose sum exactly equals
the term in Eq.~(\ref{eqGfexp}) multiplying the
$\tilde{\boldsymbol{\rho}}_1\cdot\tilde{\boldsymbol{\rho}}_2$.

There are important features of  Eq.~(\ref{eqGamHO}) that affect how
the scattering information is extracted.
For example, the
sums and integrals that correspond to the open channels all contain
$\sin (\tilde{k}_\alpha\Delta\tilde{z})$
and therefore they do not contribute to $\Gamma$ as $\Delta\tilde{z}\to 0^+$ for $\Delta \rho \ll 1$.
Terms which refer to the closed channels do
survive in this limit and they contribute to the $\eta_{\ell m}(E)$.
Lines 4 and 6 contribute to the $\eta_{1\pm 1}$ while the second
term in each of lines 1 and 2 contribute to the $\eta_{00}$ and $\eta_{10}$.
The $\eta_{00}$ is the limit as all spatial coordinates of
$\Gamma$ go to 0. The $\eta_{10}=-3\partial^2\Gamma /\partial
(\Delta\tilde{z})^2$ in the limit all spatial coordinates go to 0. This gives
\begin{eqnarray}\label{eqEtaHOapp}
\eta_{00}&=&2\sum_{n=n_{o,0}+1}^\infty\frac{ e^{-\tilde{\kappa}_{n0}
\Delta\tilde{z}}}{\tilde{\kappa}_{n0}}-2\int_{n_{o,0}}^\infty
\frac{ e^{-\tilde{\kappa}_{n0}\Delta\tilde{z}}}{
\tilde{\kappa}_{n0}}dn\cr
\eta_{10}&=&-6\sum_{n=n_{o,0}+1}^\infty\tilde{\kappa}_{n0}
e^{-\tilde{\kappa}_{n0}
\Delta\tilde{z}}+6\int_{n_{o,0}}^\infty \tilde{\kappa}_{n0}
e^{-\tilde{\kappa}_{n0}
\Delta\tilde{z}}dn\cr
\eta_{1\pm 1}&=&12\sum_{n=n_{o,1}+1}^\infty
\frac{(n+1)e^{-\tilde{\kappa}_{n1}\Delta\tilde{z}}}{\tilde{\kappa}_{n1}}\cr
&\null&-12\int_{n_{o,1}}^\infty
\frac{(n+1)e^{-\tilde{\kappa}_{n1}\Delta\tilde{z}}}{\tilde{\kappa}_{n1}}dn
\end{eqnarray}
where the limit $\Delta\tilde{z}\to 0^+$ is understood for all expressions.
Notice that all of the terms have the form of a sum
minus a corresponding integral.
This feature ensures that the $\eta_{\ell m}$ are well defined for any
finite $\Delta\tilde{z}$ and thus they give unambiguous, finite results
in the limit $\Delta\tilde{z}\to 0^+$.

To summarize, the expressions needed for the calculation of the $K$-matrix
are convergent series when using a finite $\Delta\tilde{z}$ in the
expression for the Green's function. When the scattering potential
has a short range, analytic expressions can be obtained in the limit
$\Delta\tilde{z}\to 0^+$.

\subsection{Infinite Square Well Confinement}\label{secGSW}

In this section, the motion of the particle is bounded in the $x$-
and $y$-direction by an infinitely square well.
The square well possesses the following form $V(x,y)=0$ when
$|x|<a_\perp /2$ {\it and} $|y|<a_\perp /2$ and $V(x,y)=\infty$ elsewhere.
As in the case of the harmonic oscillator at large distances the
wave function of the electron separates in Cartesian coordinates,
where the corresponding solutions of the Hamiltonian $\hat{H}_c$
possess the form in Eq.~(\ref{eqPsi0eo}) for even or odd parity in the $z$-direction,
respectively.
In addition, the energy spectrum of the $\hat{H}_c$ Hamiltonian
obeys the following relation $E=E_{n_x,n_y}+\hbar^2k^2_{n_x,n_y}/(2\mu )$
where $E_{n_xn_y}=\frac{\hbar^2\pi^2}{2\mu a_\perp^2}[n_x^2
+ n_y^2]$. For the $m=0$ case, we are only interested in the values of the quantum numbers $n$ that
give non-zero functions at the origin; these correspond to the cases
$n_x,n_y=(2\alpha_x+1),(2\alpha_y+1)$ where the $\alpha_x,\alpha_y$ are
independently $0,1,2,...$. For the $m=1$ case, we use the functions
\begin{equation}\label{eqlincomp}
\Phi^\pm_{n_xn_y}(x,y)=\frac{1}{\sqrt{2}}[F_{n_x}(x)F_{n_y}(y)\pm
i F_{n_x}(y)F_{n_y}(x)]
\end{equation}
where $F_n(x)=\sqrt{2/a_\perp}\sin (n\pi [\tilde{x}-1/2])$ with
$\tilde{x}=x/a_\perp$ and
$n_x,n_y=(2\alpha_x+2),(2\alpha_y+1)$ with the $\alpha_x,\alpha_y$
independently $0,1,2,...$. This choice gives $F'_{n_x}(0)\neq 0$ and
$F_{n_y}(0)\neq 0$.
This particular linear combination in Eq.~(\ref{eqlincomp}) ensures that
collisions with $m=1$ affect the square well states whose
$x$-dependent factor vanishes at the origin while the $y$-dependent factor is nonzero, or vice versa.
Evaluation of the square well eigensolutions at the origin  gives
\begin{equation}
  \begin{cases}
    \Phi_{\alpha_x\alpha_y}(0,0) = \frac{2(-1)^{\alpha_x+\alpha_y}}{a_\perp}~~{\rm{for}}~m=0\\
    (\partial_\pm \Phi^\pm_{\alpha_x\alpha_y})\big|_{0}=
    \frac{\sqrt{2}(-1)^{\alpha_x+\alpha_y+1}}{a_\perp}
    \frac{2\pi (\alpha_x+1)}{a_\perp}~~{\rm{for}}~m=1,
  \end{cases}
  \label{eqswphi}
 \end{equation}
where the functions $\Phi^\pm_{n_xn_y}(x,y)$ are defined in
Eq.~(\ref{eqlincomp})

We define the scaled energy to be
$\epsilon =E/(4\hbar^2\pi^2/[2\mu a_\perp^2])$. The scaled momenta can be
written as $\tilde{k}=2\pi\sqrt{\epsilon}$, $\tilde{k}_{\alpha m}=
2\pi \sqrt{\epsilon - [(\alpha_x+[|m|+1]/2)^2+(\alpha_y+1/2)^2]}$, and
$\tilde{\kappa}_{\alpha m}=
2\pi \sqrt{(\alpha_x+[|m|+1]/2)^2+(\alpha_y+1/2)^2-\epsilon}$.
Substituting Eq.~(\ref{eqswphi}) in Eq.~(\ref{eqGamma}) the following
relation is obtained:
\small
\begin{eqnarray}\label{eqGamSW}
\Gamma(\boldsymbol{r}_1,\boldsymbol{r}_2,E) =
&-&8\pi\sum_{\alpha=0}^{\alpha_{o,0}}\frac{\sin (\tilde{k}_{\alpha 0}
\Delta\tilde{z})}{\tilde{k}_{\alpha 0}}
+8\pi\sum_{\alpha=\alpha_{o,0}+1}^\infty\frac{ e^{-\tilde{\kappa}_{\alpha 0}
\Delta\tilde{z}}}{\tilde{\kappa}_{\alpha 0}}\cr
&+&8\pi\int_{-1/2}^{\alpha_{o,0}}\frac{\sin (\tilde{k}_{\alpha 0}
\Delta\tilde{z})}{\tilde{k}_{\alpha 0}} d^2\alpha
-8\pi\int_{\alpha_{o,0}}^\infty \frac{ e^{-\tilde{\kappa}_{\alpha0}
\Delta\tilde{z}}}{
\tilde{\kappa}_{\alpha 0}}d^2\alpha\cr
&-&\tilde{\boldsymbol{\rho}}_1\cdot\tilde{\boldsymbol{\rho}}_2
32\pi^3\sum_{\alpha=0}^{\alpha_{o,1}}\frac{(\alpha_x+1)^2
\sin (\tilde{k}_{\alpha 1}\Delta\tilde{z})}{
\tilde{k}_{\alpha 1}}
\cr
&+&\tilde{\boldsymbol{\rho}}_1\cdot\tilde{\boldsymbol{\rho}}_2
32\pi^3\int_{-1}^{\alpha_{o,1}}\frac{(\alpha_x+1)^2
\sin (\tilde{k}_{\alpha 1}\Delta\tilde{z})}{
\tilde{k}_{\alpha 1}}d^2\alpha
\cr
&+&\tilde{\boldsymbol{\rho}}_1\cdot\tilde{\boldsymbol{\rho}}_2
32\pi^3\sum_{\alpha=\alpha_{o,1}+1}^\infty
\frac{(\alpha_x+1)^2e^{-\tilde{\kappa}_{\alpha 1}
\Delta\tilde{z}}}{\tilde{\kappa}_{\alpha 1}}\cr
&-&\tilde{\boldsymbol{\rho}}_1\cdot\tilde{\boldsymbol{\rho}}_2
32\pi^3\int_{\alpha_{o,1}}^\infty
\frac{(\alpha_x+1)^2e^{-\tilde{\kappa}_{\alpha 1}
\Delta\tilde{z}}}{\tilde{\kappa}_{\alpha 1}}d^2\alpha
\end{eqnarray}
\normalsize
where $\alpha_{o,m}$ denotes the divide between open and closed
channels and is given by
$(\alpha_x+[|m|+1]/2)^2+(\alpha_y+1/2)^2=\epsilon$;
in the sums, the integer part of $\alpha_x,\alpha_y$ should be used. 
Note that $\tilde{\boldsymbol{\rho}}_1 \cdot \tilde{\boldsymbol{\rho}}_2=\boldsymbol{\rho}_1 \cdot \boldsymbol{\rho}_2/a_\perp^2$.
This expression has many terms but each can be identified with different
parts of the Eq.~(\ref{eqGamma}). The first line is the contribution
from the $m=0$ open (first term) and closed (second term) channels
from $\hat{G}_c$. The second line is an integral expression that exactly
equals $-\cos(\tilde{k}\Delta\tilde{z})/\Delta\tilde{z}$
(see Eq.~(\ref{eqGfexp})) which is the
contribution to $m=0$ from $G_f$; the lower limit of integration is
$-1/2$ for both $\alpha_x$ and $\alpha_y$.
The third (fifth) line is the contribution
to $m=\pm 1$ from the open (closed) channels in $\hat{G}_c$. The fourth
and sixth lines are integrals whose sum exactly equals
the term in Eq.~(\ref{eqGfexp}) multiplying the
$\tilde{\boldsymbol{\rho}}_1\cdot\tilde{\boldsymbol{\rho}}_2$;
the lower limit of integration is $-1$ for $\alpha_x$ and $-1/2$ for
$\alpha_y$. Unlike the integral expressions for the harmonic oscillator,
it is somewhat tricky to show that the integrals exactly equal the
expressions from $G_f$. The simplest method we found was to change
variables $s_x =\alpha_x +(|m|+1)/2$ and $s_y=\alpha_y+1/2$ and then
convert to cylindrical coordinates $s^2=s_x^2+s_y^2$ and $\tan (\phi )=y/x$
and remember to only integrate from $0\leq \phi\leq \pi /2$ since the
$s_x\geq 0$ and $s_y\geq 0^+$.

To obtain the scattering parameters $\eta_{\ell m}$, we can repeat the
logic of the preceding section for the harmonic oscillator.
This gives
\small
\begin{eqnarray}\label{eqEtaSWapp}
\eta_{00}&=&8\pi\sum_{\alpha_{o,0}+1}^\infty
\frac{ e^{-\tilde{\kappa}_{\alpha 0}
\Delta\tilde{z}}}{\tilde{\kappa}_{\alpha 0}}-8\pi\int_{\alpha_{o,0}}^\infty
\frac{ e^{-\tilde{\kappa}_{\alpha 0}\Delta\tilde{z}}}{
\tilde{\kappa}_{\alpha 0}}d^2\alpha\cr
\eta_{10}&=&-24\pi \sum_{\alpha_{o,0}+1}^\infty\tilde{\kappa}_{\alpha 0}
e^{-\tilde{\kappa}_{\alpha 0}
\Delta\tilde{z}}+24\pi\int_{\alpha_{o,0}}^\infty \tilde{\kappa}_{\alpha 0}
e^{-\tilde{\kappa}_{\alpha 0}
\Delta\tilde{z}}d^2\alpha \cr
\eta_{1\pm 1}&=&96\pi^3\sum_{\alpha_{o,1}+1}^\infty
\frac{(\alpha_x+1)^2e^{-\tilde{\kappa}_{\alpha 1}
\Delta\tilde{z}}}{\tilde{\kappa}_{\alpha 1}}\cr
&\null&-96\pi^3\int_{\alpha_{o,1}}^\infty
\frac{(\alpha_x+1)^2e^{-\tilde{\kappa}_{\alpha 1}
\Delta\tilde{z}}}{\tilde{\kappa}_{\alpha 1}}d^2\alpha
\end{eqnarray}
\normalsize
where the limit $\Delta\tilde{z}\to 0^+$ is understood for all expressions.
Notice that all of the terms have the form of a sum
minus a corresponding integral.
This feature ensures that the $\eta_{\ell m}$ are well defined for any
finite $\Delta\tilde{z}$ and thus they give unambiguous, finite results
in the limit $\Delta\tilde{z}\to 0$.

To summarize, the expressions needed for the calculation of the $K$-matrix
are convergent series when using a finite $\Delta\tilde{z}$ in the
expression for the Green's function. When the scattering potential
is short range, analytic expressions can be obtained in the limit
$\Delta\tilde{z}\to 0^+$.

\bibliography{atom_lib}

\begin{thebibliography}{36}%
\makeatletter
\providecommand \@ifxundefined [1]{%
 \@ifx{#1\undefined}
}%
\providecommand \@ifnum [1]{%
 \ifnum #1\expandafter \@firstoftwo
 \else \expandafter \@secondoftwo
 \fi
}%
\providecommand \@ifx [1]{%
 \ifx #1\expandafter \@firstoftwo
 \else \expandafter \@secondoftwo
 \fi
}%
\providecommand \natexlab [1]{#1}%
\providecommand \enquote  [1]{``#1''}%
\providecommand \bibnamefont  [1]{#1}%
\providecommand \bibfnamefont [1]{#1}%
\providecommand \citenamefont [1]{#1}%
\providecommand \href@noop [0]{\@secondoftwo}%
\providecommand \href [0]{\begingroup \@sanitize@url \@href}%
\providecommand \@href[1]{\@@startlink{#1}\@@href}%
\providecommand \@@href[1]{\endgroup#1\@@endlink}%
\providecommand \@sanitize@url [0]{\catcode `\\12\catcode `\$12\catcode
  `\&12\catcode `\#12\catcode `\^12\catcode `\_12\catcode `\%12\relax}%
\providecommand \@@startlink[1]{}%
\providecommand \@@endlink[0]{}%
\providecommand \url  [0]{\begingroup\@sanitize@url \@url }%
\providecommand \@url [1]{\endgroup\@href {#1}{\urlprefix }}%
\providecommand \urlprefix  [0]{URL }%
\providecommand \Eprint [0]{\href }%
\providecommand \doibase [0]{http://dx.doi.org/}%
\providecommand \selectlanguage [0]{\@gobble}%
\providecommand \bibinfo  [0]{\@secondoftwo}%
\providecommand \bibfield  [0]{\@secondoftwo}%
\providecommand \translation [1]{[#1]}%
\providecommand \BibitemOpen [0]{}%
\providecommand \bibitemStop [0]{}%
\providecommand \bibitemNoStop [0]{.\EOS\space}%
\providecommand \EOS [0]{\spacefactor3000\relax}%
\providecommand \BibitemShut  [1]{\csname bibitem#1\endcsname}%
\let\auto@bib@innerbib\@empty
\bibitem [{\citenamefont {Fano}(1981)}]{fano1981stark}%
  \BibitemOpen
  \bibfield  {author} {\bibinfo {author} {\bibfnamefont {U.}~\bibnamefont
  {Fano}},\ }\href@noop {} {\bibfield  {journal} {\bibinfo  {journal} {Phys.
  Rev. A}\ }\textbf {\bibinfo {volume} {24}},\ \bibinfo {pages} {619} (\bibinfo
  {year} {1981})}\BibitemShut {NoStop}%
\bibitem [{\citenamefont {Harmin}(1982{\natexlab{a}})}]{harminpra1982}%
  \BibitemOpen
  \bibfield  {author} {\bibinfo {author} {\bibfnamefont {D.~A.}\ \bibnamefont
  {Harmin}},\ }\href {\doibase 10.1103/PhysRevA.26.2656} {\bibfield  {journal}
  {\bibinfo  {journal} {Phys. Rev. A}\ }\textbf {\bibinfo {volume} {26}},\
  \bibinfo {pages} {2656} (\bibinfo {year} {1982}{\natexlab{a}})}\BibitemShut
  {NoStop}%
\bibitem [{\citenamefont {Harmin}(1982{\natexlab{b}})}]{harminprl1982}%
  \BibitemOpen
  \bibfield  {author} {\bibinfo {author} {\bibfnamefont {D.~A.}\ \bibnamefont
  {Harmin}},\ }\href {\doibase 10.1103/PhysRevLett.49.128} {\bibfield
  {journal} {\bibinfo  {journal} {Phys. Rev. Lett.}\ }\textbf {\bibinfo
  {volume} {49}},\ \bibinfo {pages} {128} (\bibinfo {year}
  {1982}{\natexlab{b}})}\BibitemShut {NoStop}%
\bibitem [{\citenamefont {Harmin}(1981)}]{harmin1981hydrogenic}%
  \BibitemOpen
  \bibfield  {author} {\bibinfo {author} {\bibfnamefont {D.~A.}\ \bibnamefont
  {Harmin}},\ }\href@noop {} {\bibfield  {journal} {\bibinfo  {journal} {Phys.
  Rev. A}\ }\textbf {\bibinfo {volume} {24}},\ \bibinfo {pages} {2491}
  (\bibinfo {year} {1981})}\BibitemShut {NoStop}%
\bibitem [{\citenamefont {Greene}(1987)}]{Greene1987pra}%
  \BibitemOpen
  \bibfield  {author} {\bibinfo {author} {\bibfnamefont {C.~H.}\ \bibnamefont
  {Greene}},\ }\href {\doibase 10.1103/PhysRevA.36.4236} {\bibfield  {journal}
  {\bibinfo  {journal} {Phys. Rev. A}\ }\textbf {\bibinfo {volume} {36}},\
  \bibinfo {pages} {4236} (\bibinfo {year} {1987})}\BibitemShut {NoStop}%
\bibitem [{\citenamefont {Wong}\ \emph {et~al.}(1988)\citenamefont {Wong},
  \citenamefont {Rau},\ and\ \citenamefont {Greene}}]{WongRauGreene1988pra}%
  \BibitemOpen
  \bibfield  {author} {\bibinfo {author} {\bibfnamefont {H.~Y.}\ \bibnamefont
  {Wong}}, \bibinfo {author} {\bibfnamefont {A.~R.~P.}\ \bibnamefont {Rau}}, \
  and\ \bibinfo {author} {\bibfnamefont {C.~H.}\ \bibnamefont {Greene}},\
  }\href {\doibase 10.1103/PhysRevA.37.2393} {\bibfield  {journal} {\bibinfo
  {journal} {Phys. Rev. A}\ }\textbf {\bibinfo {volume} {37}},\ \bibinfo
  {pages} {2393} (\bibinfo {year} {1988})}\BibitemShut {NoStop}%
\bibitem [{\citenamefont {Rau}\ and\ \citenamefont
  {Wong}(1988)}]{RauWong1988pra}%
  \BibitemOpen
  \bibfield  {author} {\bibinfo {author} {\bibfnamefont {A.~R.~P.}\
  \bibnamefont {Rau}}\ and\ \bibinfo {author} {\bibfnamefont {H.~Y.}\
  \bibnamefont {Wong}},\ }\href {\doibase 10.1103/PhysRevA.37.632} {\bibfield
  {journal} {\bibinfo  {journal} {Phys. Rev. A}\ }\textbf {\bibinfo {volume}
  {37}},\ \bibinfo {pages} {632} (\bibinfo {year} {1988})}\BibitemShut
  {NoStop}%
\bibitem [{\citenamefont {Greene}\ and\ \citenamefont
  {Rouze}(1988)}]{GreeneRouze1988ZPhys}%
  \BibitemOpen
  \bibfield  {author} {\bibinfo {author} {\bibfnamefont {C.~H.}\ \bibnamefont
  {Greene}}\ and\ \bibinfo {author} {\bibfnamefont {N.}~\bibnamefont {Rouze}},\
  }\href@noop {} {\bibfield  {journal} {\bibinfo  {journal} {Z. Phys. D-Atoms,
  Molecules and Clusters}\ }\textbf {\bibinfo {volume} {9}},\ \bibinfo {pages}
  {219} (\bibinfo {year} {1988})}\BibitemShut {NoStop}%
\bibitem [{\citenamefont {Slonim}\ and\ \citenamefont
  {Greene}(1991)}]{SlonimGreene1991}%
  \BibitemOpen
  \bibfield  {author} {\bibinfo {author} {\bibfnamefont {V.~Z.}\ \bibnamefont
  {Slonim}}\ and\ \bibinfo {author} {\bibfnamefont {C.~H.}\ \bibnamefont
  {Greene}},\ }\href {\doibase 10.1080/10420159108211499} {\bibfield  {journal}
  {\bibinfo  {journal} {Radiation effects and defects in solids}\ }\textbf
  {\bibinfo {volume} {122}},\ \bibinfo {pages} {679} (\bibinfo {year}
  {1991})}\BibitemShut {NoStop}%
\bibitem [{\citenamefont {Demkov}\ and\ \citenamefont
  {Drukarev}(1966)}]{demkovjetp1966}%
  \BibitemOpen
  \bibfield  {author} {\bibinfo {author} {\bibfnamefont {Y.~N.}\ \bibnamefont
  {Demkov}}\ and\ \bibinfo {author} {\bibfnamefont {G.~F.}\ \bibnamefont
  {Drukarev}},\ }\href@noop {} {\bibfield  {journal} {\bibinfo  {journal}
  {JETP}\ }\textbf {\bibinfo {volume} {2}},\ \bibinfo {pages} {182} (\bibinfo
  {year} {1966})}\BibitemShut {NoStop}%
\bibitem [{\citenamefont {Grozdanov}(1995)}]{grozdanovpra1995}%
  \BibitemOpen
  \bibfield  {author} {\bibinfo {author} {\bibfnamefont {T.~P.}\ \bibnamefont
  {Grozdanov}},\ }\href {\doibase 10.1103/PhysRevA.51.607} {\bibfield
  {journal} {\bibinfo  {journal} {Phys. Rev. A}\ }\textbf {\bibinfo {volume}
  {51}},\ \bibinfo {pages} {607} (\bibinfo {year} {1995})}\BibitemShut
  {NoStop}%
\bibitem [{\citenamefont {Olshanii}(1998)}]{olshanii1998prl}%
  \BibitemOpen
  \bibfield  {author} {\bibinfo {author} {\bibfnamefont {M.}~\bibnamefont
  {Olshanii}},\ }\href {\doibase 10.1103/PhysRevLett.81.938} {\bibfield
  {journal} {\bibinfo  {journal} {Phys. Rev. Lett.}\ }\textbf {\bibinfo
  {volume} {81}},\ \bibinfo {pages} {938} (\bibinfo {year} {1998})}\BibitemShut
  {NoStop}%
\bibitem [{\citenamefont {Granger}\ and\ \citenamefont
  {Blume}(2004)}]{GrangerBlume2004prl}%
  \BibitemOpen
  \bibfield  {author} {\bibinfo {author} {\bibfnamefont {B.~E.}\ \bibnamefont
  {Granger}}\ and\ \bibinfo {author} {\bibfnamefont {D.}~\bibnamefont
  {Blume}},\ }\href {\doibase 10.1103/PhysRevLett.92.133202} {\bibfield
  {journal} {\bibinfo  {journal} {Phys. Rev. Lett}\ }\textbf {\bibinfo {volume}
  {92}},\ \bibinfo {pages} {133202} (\bibinfo {year} {2004})}\BibitemShut
  {NoStop}%
\bibitem [{\citenamefont {Giannakeas}\ \emph {et~al.}(2012)\citenamefont
  {Giannakeas}, \citenamefont {Diakonos},\ and\ \citenamefont
  {Schmelcher}}]{Giannakeas2012pra}%
  \BibitemOpen
  \bibfield  {author} {\bibinfo {author} {\bibfnamefont {P.}~\bibnamefont
  {Giannakeas}}, \bibinfo {author} {\bibfnamefont {F.~K.}\ \bibnamefont
  {Diakonos}}, \ and\ \bibinfo {author} {\bibfnamefont {P.}~\bibnamefont
  {Schmelcher}},\ }\href@noop {} {\bibfield  {journal} {\bibinfo  {journal}
  {Phys. Rev. A}\ }\textbf {\bibinfo {volume} {86}},\ \bibinfo {pages} {042703}
  (\bibinfo {year} {2012})}\BibitemShut {NoStop}%
\bibitem [{\citenamefont {He\ss{}}\ \emph {et~al.}(2014)\citenamefont
  {He\ss{}}, \citenamefont {Giannakeas},\ and\ \citenamefont
  {Schmelcher}}]{hess2014pra}%
  \BibitemOpen
  \bibfield  {author} {\bibinfo {author} {\bibfnamefont {B.}~\bibnamefont
  {He\ss{}}}, \bibinfo {author} {\bibfnamefont {P.}~\bibnamefont {Giannakeas}},
  \ and\ \bibinfo {author} {\bibfnamefont {P.}~\bibnamefont {Schmelcher}},\
  }\href {http://link.aps.org/doi/10.1103/PhysRevA.89.052716} {\bibfield
  {journal} {\bibinfo  {journal} {Phys. Rev. A}\ }\textbf {\bibinfo {volume}
  {89}},\ \bibinfo {pages} {052716} (\bibinfo {year} {2014})}\BibitemShut
  {NoStop}%
\bibitem [{\citenamefont {Zhang}\ and\ \citenamefont
  {Greene}(2013)}]{Zhang2013pra}%
  \BibitemOpen
  \bibfield  {author} {\bibinfo {author} {\bibfnamefont {C.}~\bibnamefont
  {Zhang}}\ and\ \bibinfo {author} {\bibfnamefont {C.~H.}\ \bibnamefont
  {Greene}},\ }\href@noop {} {\bibfield  {journal} {\bibinfo  {journal} {Phys.
  Rev. A}\ }\textbf {\bibinfo {volume} {88}},\ \bibinfo {pages} {012715}
  (\bibinfo {year} {2013})}\BibitemShut {NoStop}%
\bibitem [{\citenamefont {Giannakeas}\ \emph {et~al.}(2013)\citenamefont
  {Giannakeas}, \citenamefont {Melezhik},\ and\ \citenamefont
  {Schmelcher}}]{giannakeas2013prl}%
  \BibitemOpen
  \bibfield  {author} {\bibinfo {author} {\bibfnamefont {P.}~\bibnamefont
  {Giannakeas}}, \bibinfo {author} {\bibfnamefont {V.~S.}\ \bibnamefont
  {Melezhik}}, \ and\ \bibinfo {author} {\bibfnamefont {P.}~\bibnamefont
  {Schmelcher}},\ }\href
  {http://link.aps.org/doi/10.1103/PhysRevLett.111.183201} {\bibfield
  {journal} {\bibinfo  {journal} {Phys. Rev. Lett.}\ }\textbf {\bibinfo
  {volume} {111}},\ \bibinfo {pages} {183201} (\bibinfo {year}
  {2013})}\BibitemShut {NoStop}%
\bibitem [{\citenamefont {Kim}\ \emph {et~al.}(2005)\citenamefont {Kim},
  \citenamefont {Schmiedmayer},\ and\ \citenamefont {Schmelcher}}]{kim2005}%
  \BibitemOpen
  \bibfield  {author} {\bibinfo {author} {\bibfnamefont {J.~I.}\ \bibnamefont
  {Kim}}, \bibinfo {author} {\bibfnamefont {J.}~\bibnamefont {Schmiedmayer}}, \
  and\ \bibinfo {author} {\bibfnamefont {P.}~\bibnamefont {Schmelcher}},\
  }\href {\doibase 10.1103/PhysRevA.72.042711} {\bibfield  {journal} {\bibinfo
  {journal} {Phys. Rev. A}\ }\textbf {\bibinfo {volume} {72}},\ \bibinfo
  {pages} {042711} (\bibinfo {year} {2005})}\BibitemShut {NoStop}%
\bibitem [{\citenamefont {Crawford}(1988)}]{crawfordpra1988}%
  \BibitemOpen
  \bibfield  {author} {\bibinfo {author} {\bibfnamefont {O.~H.}\ \bibnamefont
  {Crawford}},\ }\href {\doibase 10.1103/PhysRevA.37.2432} {\bibfield
  {journal} {\bibinfo  {journal} {Phys. Rev. A}\ }\textbf {\bibinfo {volume}
  {37}},\ \bibinfo {pages} {2432} (\bibinfo {year} {1988})}\BibitemShut
  {NoStop}%
\bibitem [{\citenamefont {Du}\ and\ \citenamefont {Greene}(1987)}]{dupra1987}%
  \BibitemOpen
  \bibfield  {author} {\bibinfo {author} {\bibfnamefont {N.}~\bibnamefont
  {Du}}\ and\ \bibinfo {author} {\bibfnamefont {C.}~\bibnamefont {Greene}},\
  }\href@noop {} {\bibfield  {journal} {\bibinfo  {journal} {Phys. Rev. A}\
  }\textbf {\bibinfo {volume} {36}},\ \bibinfo {pages} {971} (\bibinfo {year}
  {1987})}\BibitemShut {NoStop}%
\bibitem [{\citenamefont {Schwinger}(1947)}]{schwinger1947}%
  \BibitemOpen
  \bibfield  {author} {\bibinfo {author} {\bibfnamefont {J.}~\bibnamefont
  {Schwinger}},\ }\href@noop {} {\bibfield  {journal} {\bibinfo  {journal}
  {Phys. Rev.}\ }\textbf {\bibinfo {volume} {72}} (\bibinfo {year}
  {1947})}\BibitemShut {NoStop}%
\bibitem [{\citenamefont {Schwinger}(1950)}]{schwinger1950}%
  \BibitemOpen
  \bibfield  {author} {\bibinfo {author} {\bibfnamefont {J.}~\bibnamefont
  {Schwinger}},\ }\href@noop {} {\bibfield  {journal} {\bibinfo  {journal}
  {Phys. Rev.}\ }\textbf {\bibinfo {volume} {78}} (\bibinfo {year}
  {1950})}\BibitemShut {NoStop}%
\bibitem [{\citenamefont {Lippmann}\ and\ \citenamefont
  {Schwinger}(1950)}]{schwingerlippmann1950}%
  \BibitemOpen
  \bibfield  {author} {\bibinfo {author} {\bibfnamefont {B.~A.}\ \bibnamefont
  {Lippmann}}\ and\ \bibinfo {author} {\bibfnamefont {J.}~\bibnamefont
  {Schwinger}},\ }\href {\doibase 10.1103/PhysRev.79.469} {\bibfield  {journal}
  {\bibinfo  {journal} {Phys. Rev.}\ }\textbf {\bibinfo {volume} {79}},\
  \bibinfo {pages} {469} (\bibinfo {year} {1950})}\BibitemShut {NoStop}%
\bibitem [{\citenamefont {Stephens}\ and\ \citenamefont
  {McKoy}(1992)}]{stephens1992jcp}%
  \BibitemOpen
  \bibfield  {author} {\bibinfo {author} {\bibfnamefont {J.~A.}\ \bibnamefont
  {Stephens}}\ and\ \bibinfo {author} {\bibfnamefont {V.}~\bibnamefont
  {McKoy}},\ }\href {\doibase 10.1063/1.463428} {\bibfield  {journal} {\bibinfo
   {journal} {The Journal of Chemical Physics}\ }\textbf {\bibinfo {volume}
  {97}},\ \bibinfo {pages} {8060} (\bibinfo {year} {1992})}\BibitemShut
  {NoStop}%
\bibitem [{\citenamefont {Goforth}\ and\ \citenamefont
  {Watson}(1992)}]{goforthpra1992}%
  \BibitemOpen
  \bibfield  {author} {\bibinfo {author} {\bibfnamefont {T.~L.}\ \bibnamefont
  {Goforth}}\ and\ \bibinfo {author} {\bibfnamefont {D.~K.}\ \bibnamefont
  {Watson}},\ }\href {\doibase 10.1103/PhysRevA.46.1239} {\bibfield  {journal}
  {\bibinfo  {journal} {Phys. Rev. A}\ }\textbf {\bibinfo {volume} {46}},\
  \bibinfo {pages} {1239} (\bibinfo {year} {1992})}\BibitemShut {NoStop}%
\bibitem [{\citenamefont {Lucchese}\ and\ \citenamefont
  {McKoy}(1979)}]{lucchese1979jpb}%
  \BibitemOpen
  \bibfield  {author} {\bibinfo {author} {\bibfnamefont {R.~R.}\ \bibnamefont
  {Lucchese}}\ and\ \bibinfo {author} {\bibfnamefont {V.}~\bibnamefont
  {McKoy}},\ }\href {\doibase 10.1088/0022-3700/12/14/005} {\bibfield
  {journal} {\bibinfo  {journal} {J. Phys. B: At. Mol. Phys.}\ }\textbf
  {\bibinfo {volume} {12}},\ \bibinfo {pages} {L421} (\bibinfo {year}
  {1979})}\BibitemShut {NoStop}%
\bibitem [{\citenamefont {Maleki}\ and\ \citenamefont
  {Macek}(1980)}]{malekipra1980}%
  \BibitemOpen
  \bibfield  {author} {\bibinfo {author} {\bibfnamefont {N.}~\bibnamefont
  {Maleki}}\ and\ \bibinfo {author} {\bibfnamefont {J.}~\bibnamefont {Macek}},\
  }\href {\doibase 10.1103/PhysRevA.21.1403} {\bibfield  {journal} {\bibinfo
  {journal} {Phys. Rev. A}\ }\textbf {\bibinfo {volume} {21}},\ \bibinfo
  {pages} {1403} (\bibinfo {year} {1980})}\BibitemShut {NoStop}%
\bibitem [{\citenamefont {Lucchese}\ and\ \citenamefont
  {McKoy}(1983)}]{lucchese1983pra}%
  \BibitemOpen
  \bibfield  {author} {\bibinfo {author} {\bibfnamefont {R.}~\bibnamefont
  {Lucchese}}\ and\ \bibinfo {author} {\bibfnamefont {V.}~\bibnamefont
  {McKoy}},\ }\href {\doibase 10.1103/PhysRevA.28.1382} {\bibfield  {journal}
  {\bibinfo  {journal} {Phys. Rev. A}\ }\textbf {\bibinfo {volume} {28}},\
  \bibinfo {pages} {1382} (\bibinfo {year} {1983})}\BibitemShut {NoStop}%
\bibitem [{\citenamefont {Lino}\ and\ \citenamefont
  {Lima}(2000)}]{lino2000bjp}%
  \BibitemOpen
  \bibfield  {author} {\bibinfo {author} {\bibfnamefont {J.~L.~S.}\
  \bibnamefont {Lino}}\ and\ \bibinfo {author} {\bibfnamefont {M.~A.~P.}\
  \bibnamefont {Lima}},\ }\href {\doibase 10.1590/S0103-97332000000200028}
  {\bibfield  {journal} {\bibinfo  {journal} {Brazilian Journal of Physics}\
  }\textbf {\bibinfo {volume} {30}},\ \bibinfo {pages} {432} (\bibinfo {year}
  {2000})}\BibitemShut {NoStop}%
\bibitem [{\citenamefont {Lucchese}\ \emph {et~al.}(1983)\citenamefont
  {Lucchese}, \citenamefont {Takatsuka}, \citenamefont {Watson},\ and\
  \citenamefont {McKoy}}]{lucchese1983amopp}%
  \BibitemOpen
  \bibfield  {author} {\bibinfo {author} {\bibfnamefont {R.~R.}\ \bibnamefont
  {Lucchese}}, \bibinfo {author} {\bibfnamefont {K.}~\bibnamefont {Takatsuka}},
  \bibinfo {author} {\bibfnamefont {D.~K.}\ \bibnamefont {Watson}}, \ and\
  \bibinfo {author} {\bibfnamefont {V.}~\bibnamefont {McKoy}}\ }(\bibinfo
  {publisher} {Springer US},\ \bibinfo {year} {1983})\ pp.\ \bibinfo {pages}
  {29--49}\BibitemShut {NoStop}%
\bibitem [{\citenamefont {Larson}\ and\ \citenamefont
  {Stoneman}(1985)}]{larsonpra1985}%
  \BibitemOpen
  \bibfield  {author} {\bibinfo {author} {\bibfnamefont {D.~J.}\ \bibnamefont
  {Larson}}\ and\ \bibinfo {author} {\bibfnamefont {R.}~\bibnamefont
  {Stoneman}},\ }\href {\doibase 10.1103/PhysRevA.31.2210} {\bibfield
  {journal} {\bibinfo  {journal} {Phys. Rev. A}\ }\textbf {\bibinfo {volume}
  {31}},\ \bibinfo {pages} {2210} (\bibinfo {year} {1985})}\BibitemShut
  {NoStop}%
\bibitem [{\citenamefont {Krause}(1990)}]{krausseprl1990}%
  \BibitemOpen
  \bibfield  {author} {\bibinfo {author} {\bibfnamefont {H.~F.}\ \bibnamefont
  {Krause}},\ }\href {\doibase 10.1103/PhysRevLett.64.1725} {\bibfield
  {journal} {\bibinfo  {journal} {Phys. Rev. Lett.}\ }\textbf {\bibinfo
  {volume} {64}},\ \bibinfo {pages} {1725} (\bibinfo {year}
  {1990})}\BibitemShut {NoStop}%
\bibitem [{\citenamefont {Blumberg}\ \emph {et~al.}(1979)\citenamefont
  {Blumberg}, \citenamefont {Itano},\ and\ \citenamefont
  {Larson}}]{larsonpra1979}%
  \BibitemOpen
  \bibfield  {author} {\bibinfo {author} {\bibfnamefont {W.~A.~M.}\
  \bibnamefont {Blumberg}}, \bibinfo {author} {\bibfnamefont {W.~M.}\
  \bibnamefont {Itano}}, \ and\ \bibinfo {author} {\bibfnamefont {D.~J.}\
  \bibnamefont {Larson}},\ }\href {\doibase 10.1103/PhysRevA.19.139} {\bibfield
   {journal} {\bibinfo  {journal} {Phys. Rev. A}\ }\textbf {\bibinfo {volume}
  {19}},\ \bibinfo {pages} {139} (\bibinfo {year} {1979})}\BibitemShut
  {NoStop}%
\bibitem [{gol(1984)}]{golubkovjphysb1984}%
  \BibitemOpen
  \href@noop {} {\bibfield  {journal} {\bibinfo  {journal} {J. Phys. B: At.
  Mol. Phys.}\ }\textbf {\bibinfo {volume} {17}},\ \bibinfo {pages} {747}
  (\bibinfo {year} {1984})}\BibitemShut {NoStop}%
\bibitem [{\citenamefont {Hess}\ \emph {et~al.}(2015)\citenamefont {Hess},
  \citenamefont {Giannakeas},\ and\ \citenamefont {Schmelcher}}]{hess15arxiv}%
  \BibitemOpen
  \bibfield  {author} {\bibinfo {author} {\bibfnamefont {B.}~\bibnamefont
  {Hess}}, \bibinfo {author} {\bibfnamefont {P.}~\bibnamefont {Giannakeas}}, \
  and\ \bibinfo {author} {\bibfnamefont {P.}~\bibnamefont {Schmelcher}},\
  }\href@noop {} {\bibfield  {journal} {\bibinfo  {journal} {arXiv preprint
  arXiv:1503.05539}\ } (\bibinfo {year} {2015})}\BibitemShut {NoStop}%
\bibitem [{\citenamefont {Kan\'asz-Nagy}\ \emph {et~al.}(2015)\citenamefont
  {Kan\'asz-Nagy}, \citenamefont {Demler},\ and\ \citenamefont
  {Zar\'and}}]{delmerpra2015}%
  \BibitemOpen
  \bibfield  {author} {\bibinfo {author} {\bibfnamefont {M.}~\bibnamefont
  {Kan\'asz-Nagy}}, \bibinfo {author} {\bibfnamefont {E.~A.}\ \bibnamefont
  {Demler}}, \ and\ \bibinfo {author} {\bibfnamefont {G.}~\bibnamefont
  {Zar\'and}},\ }\href {\doibase 10.1103/PhysRevA.91.032704} {\bibfield
  {journal} {\bibinfo  {journal} {Phys. Rev. A}\ }\textbf {\bibinfo {volume}
  {91}},\ \bibinfo {pages} {032704} (\bibinfo {year} {2015})}\BibitemShut
  {NoStop}%
\end{thebibliography}%

\end{document}